\documentclass{jfm}
\usepackage{graphicx}
\usepackage{epstopdf, epsfig}
\usepackage{booktabs}

\newcommand{\pd}[2]{\frac{\partial #1}{\partial #2}}
\newcommand{\ave}[1]{\left \langle #1 \right \rangle}

\shorttitle{Scale interactions in turbulent plane Couette flows in minimal domains}
\shortauthor{T.~Kawata and T.~Tsukahara}

\title{Scale interactions in turbulent plane Couette flows in minimal domains}
\author{Takuya Kawata\aff{1}\corresp{\email{kawata@keio.jp}}
 \and Takahiro Tsukahara\aff{2}\corresp{\email{tsuka@rs.tus.ac.jp}}}

\affiliation{\aff{1}Department of Mechanical Engineering, Keio University, Yokohama 223-8522, Japan
\aff{2}Department of Mechanical Engineering, Tokyo University of Science, Chiba 278-8510, Japan
}

\begin{document}

\maketitle

\begin{abstract}
Interscale energy transfer in wall turbulence has been intensively studied in recent years, and both the forward (i.e.~from larger to smaller scales) and reversed transfers of turbulent energy have been found whereas their corresponding physical phenomena have not been revealed. In the present study, we perform DNS of turbulent plane Couette flow with reduced-size computational domains, where either the streamwise or spanwise domain size is reduced to their minimal lengths, aiming at elucidating the role of scale interactions in each direction. Our computational results with the streamwise-minimal domain suggest that the interplays between streamwise-elongated streaks and vortices smaller than the streamwise minimal length are the essential scale interactions for both the inner and outer structures. We further show that these streamwise-independent and smaller-scale structures exchange energy through forward and reversed interscale energy transfers, and the reversed energy transfer maintains the energy production at larger scales. Based on the resemblance of the observed Reynolds-stress transport and the scenario of the self-sustaining cycle, we conjecture that the forward and reversed energy transfers mainly represent the streak instabilities and regeneration of streamwise vortices, respectively. Furthermore, the computation with the spanwise-minimal domain indicates that the interscale energy transfers observed by one-dimensional spanwise spectral analysis are likely related to individual dynamics of each inner and outer structure, rather than represent their interactions.
\end{abstract}


\section{Introduction}\label{sec:intro}
With the significant advances in experimental techniques and high-performance computation in recent years, there have been overwhelming evidences indicating the existence of large-scale structures away from the wall in wall turbulence~\citep[see, for example,][and the references therein]{smits_2011}, and it has been observed that the influences from the large-scale outer structure affects the magnitude of small-scale events in the near-wall region~\citep[e.g.][]{hutchins_2007,dogan_2019}. Such a top-down effect from the outer to inner structure is often referred to as `amplitude modulation', and the degree of the modulation has been reported to be increasingly significant at higher Reynolds numbers~\citep[e.g.][]{mathis_2009a}. Recent experimental data from high-Reynolds-number facilities such as the Princeton Superpipe~\citep{hultmark_2012} and the CICLoPE at the University of Bologna~\citep{willert_2017,samie_2018} have clearly shown emergence of the outer peak of the streamwise velocity fluctuation at very high Reynolds numbers, indicating that the large-scale structures may play further important roles at higher Reynolds numbers including their interferences on the near-wall structures. These observations have raised the interests in energy transfer caused by interactions between different scales in wall turbulence, and recently there have been intensive efforts to investigate the interscale energy transfer~\citep[e.g.][]{cimarelli_2013,cimarelli_2016,cho_2018,hamba_2015,hamba_2018,mizuno_2016,lee_2015b,lee_2019,kawata_2018,bauer_2019,motoori_2019}. Interestingly, some studies reported not only the energy transfer from larger to smaller scales but also the reversed transfer from smaller to larger scales, whereas the corresponding physical phenomena have still not been clearly identified.  

Turbulent plane Couette flow is well known to involve a very-large-scale structure in the channel core region which fills the entire channel gap and has an extremely large streamwise extent~\cite[for example,][]{lee_1991,bech_1995,nils_phd,papavassiliou_1997,kitoh_2005,kitoh_2008,tsukahara_2006,tsukahara_2007,tsukahara_2010,avsarkisov_2014,pirozzoli_2014,kawata_2016b,lee_2018}. The uniqueness of this flow is that the very-large-scale structure appears at moderate Reynolds numbers due to the non-zero turbulent energy production at the channel centre, and therefore clear scale separation between such outer structure and smaller-scale structure near the wall can be achieved at relatively lower Reynolds numbers compared to other wall turbulence configurations, such as turbulent channels, pipes and boundary layers. For instance, the lower limit of the K\'{a}rm\'{a}n number, which is equivalent to the friction Reynolds number $Re_\tau$ used in the present study, for clear separation between the inner and outer structures in the turbulent channel flow has been reported to be about 1020~\citep[e.g.][]{monty_2009}, whereas in the turbulent plane Couette flow both the near-wall and very-large-scale structures are clearly observed already at $Re_\tau \approx 120$ as will be shown by the results of the present study in Sec.~\ref{sec:result}. Therefore, the turbulent plane Couette flow may serve as a good test case to study nonlinear interactions between coherent structures at different scales in wall turbulence. 

However, the existence of the very-large-scale structure raises an issue of computational domain size for direct numerical simulation (DNS) of the turbulent plane Couette flow as pointed out by, for example, \cite{komminaho_1996}. As the streamwise and spanwise extents of the structure have been reported as 20--30 times and 2--2.5 times of the channel height~\cite[e.g.][]{tsukahara_2006, avsarkisov_2014}, respectively, one needs to use an extremely large computational domain to exclude the domain size effect, which makes it difficult to perform the DNS of this flow at high Reynolds numbers.

Capturing all scales (or wavelengths) involved in the flow dynamics using a sufficiently large computational domain in DNS is important to, for instance, achieve a quantitative agreement with experimental data. However, the use of such a large computational domain may not be necessary when it comes to extracting and understanding the essential physics of wall turbulence. In fact, the minimal domain,  i.e.~the smallest computational domain size to maintain wall turbulence, was successfully used to substantially limit degrees of freedom of the flow and thereby extract the essential structure sustaining the wall turbulence~\citep[e.g.][]{jimenez_1991,hamilton_1995,waleffe_1997,flores_2010,hwang_2016a,hwang_2016b}. In particular, \cite{toh_2005} employed a `streamwise-minimal domain', whose spanwise extent is sufficiently large but streamwise one is minimal, for their numerical simulation of a turbulent channel flow. Their purpose of using such a domain was to simplify the dynamics of the large-scale structure in the outer layer and thereby investigate their interactions with the near-wall structure. Later, usefulness of the streamwise-minimal domain was further demonstrated by DNSs of turbulent channel~\citep{abe_2018} and pipe~\citep{han_2019} flows at relatively high Reynolds numbers ($Re_\tau \sim 1000$ or larger). 

In the present study, we perform series of DNS of turbulent plane Couette flow where either the streamwise or spanwise domain length is systematically reduced to be as small as their minimal lengths, and investigate how scale interactions in this flow are affected by such domain-size reductions. In particular, in the case of the streamwise-minimal domain we focus on why the simulation with such a reduced-size domain still can reproduce the flow features given by a sufficiently large domain. In the spanwise-minimal case, on the other hand, the very-large-scale structure of the turbulent plane Couette flow does not appear due to the too-small spanwise domain width. Our focus is therefore placed on how the turbulence transport by scale interactions is affected by the disappearance of the very-large-scale structure, aiming at elucidating the role of interactions between the inner and outer structures in the turbulence interscale transport.

\section{Numerical setup}
\label{sec:num}
We consider a plane Couette flow where one of the walls is stationary and the other is translating with a constant speed $U_\mathrm{w}$, and the distance between the walls is $h$. The $x$, $y$ and $z$-axes denote the streamwise, wall-normal and spanwise directions, and the stationary and translating walls are located at $y=0$ and $y=h$, respectively. 

\begin{table}
\begin{center}
	\begin{tabular}{rccccc}
		&	    	 $L_x$, $L_z$ ($L_x^+$, $L_z^+$) &$N_x \times N_y \times N_z$ & $\Delta x^+$, $\Delta z^+$ & $\Delta y^+$ & $Re_\tau$ \\ \midrule
		Reference case \hspace{3.5pt} & $96h$, $12.8h$ (24230, 3231) & $2048 \times 96 \times 512$ & 11.83, 6.31 & 0.18--5.66 & 126.2 \\[5pt]  
		Case A1 \hspace{3.5pt} & $24.0h$, $12.8h$ (6031, 3222) & $512 \times 96 \times512$ & 11.80, 6.29 &  0.26--6.14 & 125.8\\ 
			   2 \hspace{3.5pt} & $6.4h$, $12.8h$ (1622, 3245) & $128 \times 96 \times 512$ & 12.67 6.34 &  0.26-- 6.19 & 126.7 \\  
			   3 \hspace{3.5pt} & $1.6h$, $12.8h$ (406, 3246) & $32 \times 96 \times 512$ & 12.68    6.34 & 0.26--6.19 & 126.8 \\[5pt]
		Case B1 \hspace{3.5pt} & $96.0h$, $0.8h$ (22843, 190) & $2048 \times 96 \times 32$ & 11.15, 5.95 & 0.24--5.81 & 119.0\\ 
		            2 \hspace{3.5pt} & $96.0h$, $0.5h$ (20906, 109) &$2048 \times 96 \times 32$ & 10.20, 3.40 &  0.22--5.31 & 108.9 \\[5pt]
		Case C \hspace{3.5pt} & $0.73h$, $0.5h$ (164, 113) & $16 \times 96 \times 32$ & 10.26, 3.52 & 0.23--5.50 & 112.6 \\ 
	\end{tabular}
	\caption{Computational conditions: domain size $L_x$ and $L_z$, number of grid points $N_x$ and $N_z$, and spatial resolutions $\Delta x$ and $\Delta z$. The Reynolds number $Re_\mathrm{w}=U_\mathrm{w} h/\nu$ and time step are 8600 and $\Delta t^\ast=0.004$, respectively, in all cases. Values of the friction Reynolds number $Re_\tau=\tau (h/2)/\nu$ obtained in these cases are also given.}
	\label{tab:cond}
\end{center}
\end{table}
The governing equations for the DNS are the continuity and Navier-Stokes equations for incompressible fluid that are non-dimensionalised by $U_\mathrm{w}$ and $h$, 
with the Reynolds number defined as $Re_\mathrm{w}=U_\mathrm{w} h/\nu$ ($\nu$ is the kinetic viscosity of the fluid). The periodic boundary condition was applied to the $x$- and $z$-directions and the non-slip condition was used on the wall. Further numerical details of the present DNS codes are found in \cite{tsukahara_2006}.

We have estimated the minimal values for the streamwise and spanwise domain size to be 400 and 100 wall units, respectively, based on the earlier numerical investigations~\citep{jimenez_1991,toh_2005}. As for the sufficiently large domain size to avoid confinement effect, $L_x = 96.0h$ and $L_z=12.8h$ are used based on the earlier study by \cite{tsukahara_phd}.  In the present study we performed two series of the DNS: Cases~A and Cases~B. In the former, $L_z$ is fixed at the large enough length $L_z=12.8h$ so that some pairs of the very-large-scale vortices in the core region are captured, while $L_x$ is systematically reduced to the minimal length $L_x=1.6h$ ($L_x^+ \approx 400$). In the latter series, on the other hand, $L_x$ is extremely long ($L_x=96.0h$) while $L_z$ is on the order of the minimal length so that only the near-wall structure is captured. We also investigated an additional case for comparison, Case~C, where both $L_x$ and $L_z$ are as small as their minimal lengths. In all cases the Reynolds number is $Re_\mathrm{w}=8600$, at which the very-large-scale structure is clearly observed~\citep{tsukahara_2006}. The results of these cases are compared with the reference case, where $(L_x, L_z)=(96.0h, 12.8h)$. Further details of the computational condition for each case, such as the number of grids and the spatial resolution, are summarised in table~\ref{tab:cond}.

\begin{figure}
\begin{center}
	\includegraphics[width=0.95\hsize]{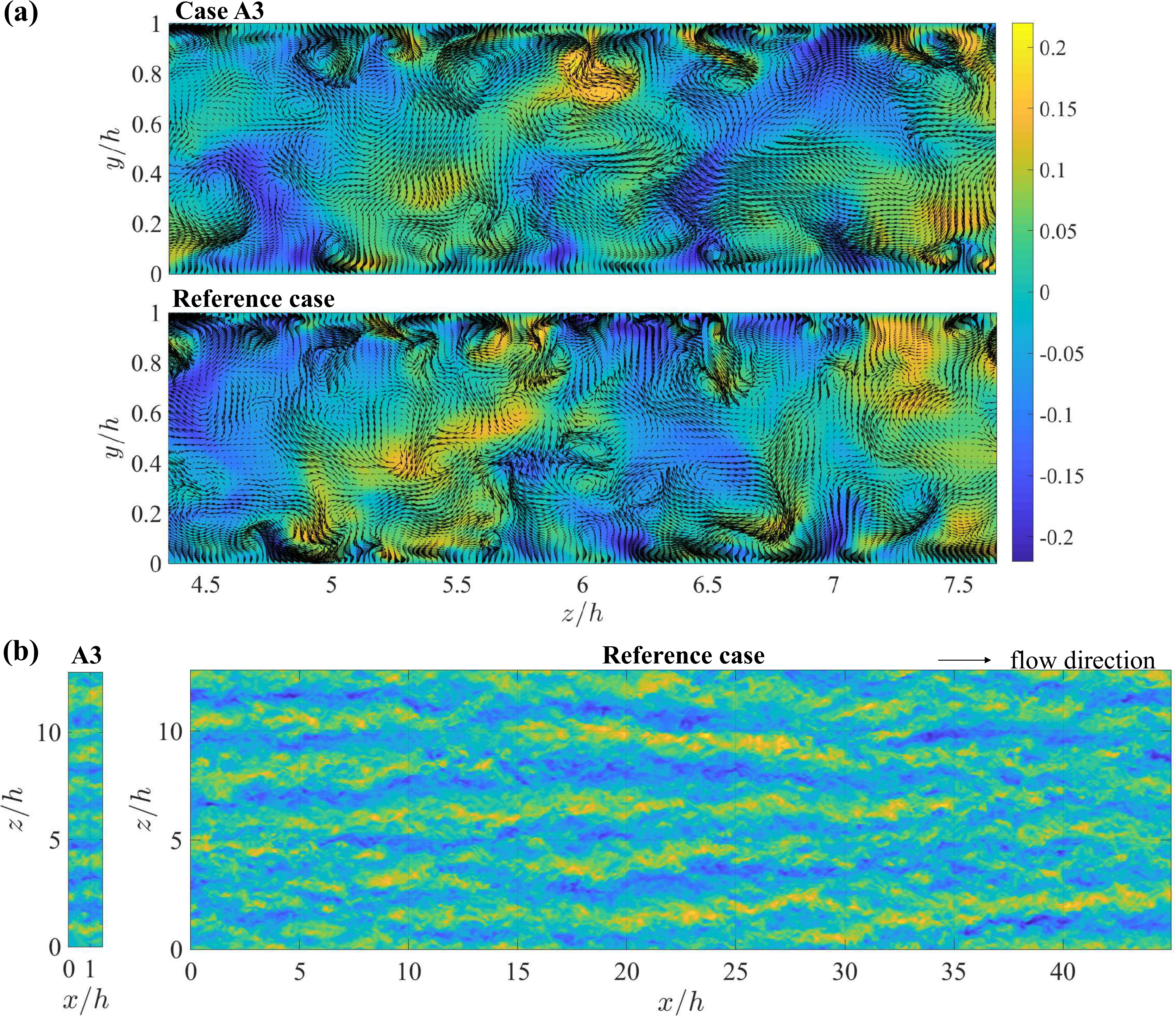}
	\caption{Snapshots of instantaneous velocity fields obtained in the A3 and reference cases on (a) streamwise cross-sectional ($zy$-) planes at arbitrary streamwise positions and (b) the channel centre plane ($xz$-plane at $y/h=0.5$). The colours in all panels represent the fluctuating streamwise velocity $u / U_\mathrm{w}$ for the range shown in the panel~(a), and the black arrows in the panel~(a) indicate the in-plane velocity vectors with the length scale where $1h$ length of the arrow corresponds to $U_\mathrm{w}$. }
	\label{fig:instAzy} 
\end{center}
\end{figure}

\section{Computational results}\label{sec:result}
\subsection{Flow structures and the basic statistics profiles}
\label{sec:flow}

In the following, the velocity components in the $x$-, $y$- and $z$-directions are denoted as $U+u$, $v$ and $w$, respectively, where $U$ is the mean streamwise velocity and the lower-case letters represent the fluctuating part of each velocity component (note that the mean values of the wall-normal and spanwise velocity components are zero). The averaged quantities given later are obtained by averaging in the $x$- and $z$-directions and in time. The instantaneous flow field obtained in the streamwise-minimal (A3) case is presented in figure~\ref{fig:instAzy} being compared with the reference-case results. As shown by the streamwise-cross sectional view in figure~\ref{fig:instAzy}(a), the streamwise-minimal case reproduces the characteristic flow features observed in the reference case quite well, exhibiting both the near-wall structures (vortical motions observed near the walls around $y/h\approx \pm 0.1$) and the very-large-scale structures (the high and low $u$ regions filling up the entire channel gap). In the $xz$-plane views at the channel centre ($y/h=0.5$) given in figure~\ref{fig:instAzy}(b), it is shown that the high- and low-$u$ streaks corresponding to the very-large-scale structures are presented in the reference case and similar streaks are found in Case~A3 as well. It is also shown that the streamwise domain extent in Case~A3 is considerably small as compared to the reference case. Due to such a reduced domain length, the spanwise meandering motions of these streaks with some splitting and/or merging, which can be clearly observed in the reference-case result, are not captured in Case~A3. 

\begin{figure}
\begin{center}
	\includegraphics[width=0.8\hsize]{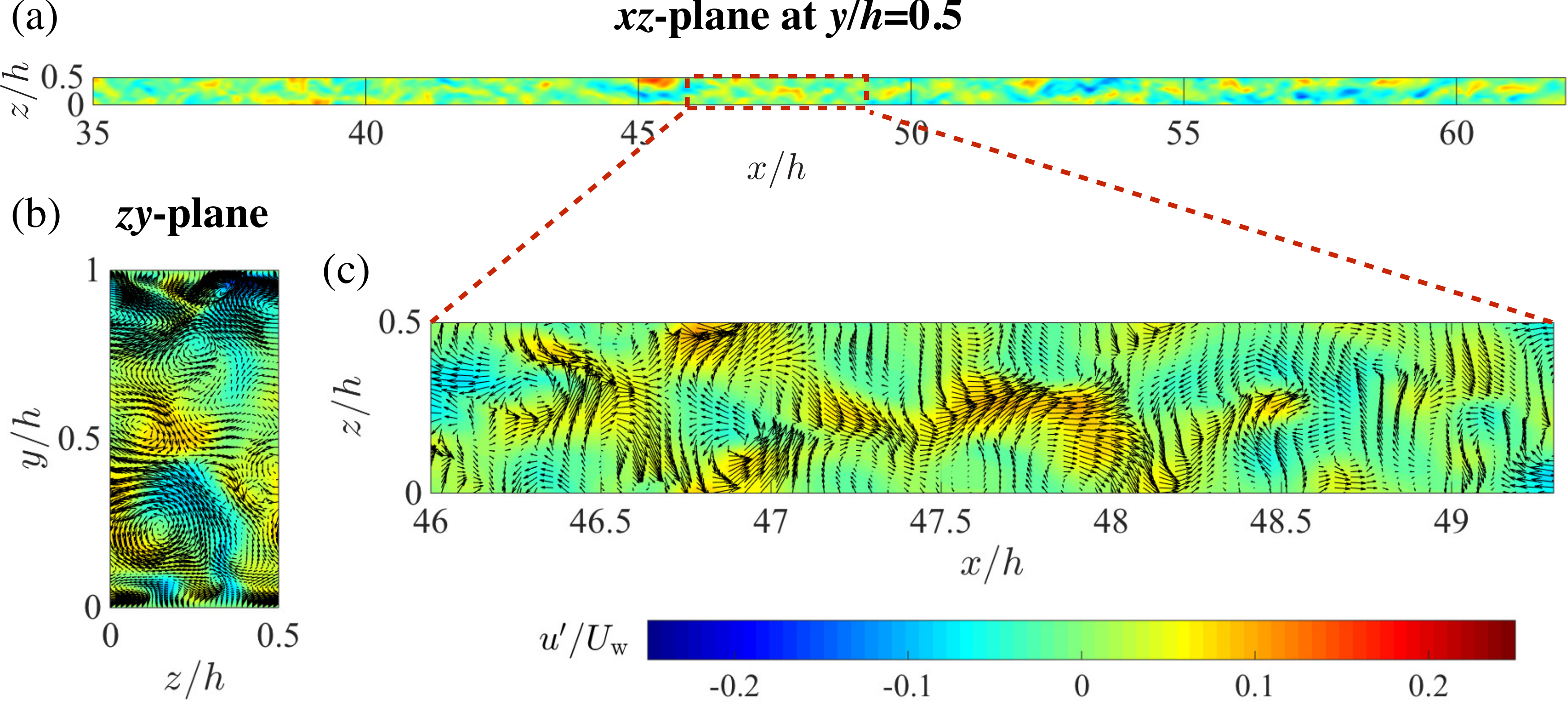}
	\caption{Snapshots of instantaneous velocity fields obtained in Case~B2: (a) a wide view of the $xz$-plane at the channel centre; (b) a streamwise cross-sectional view at $x=0.5L_x$ ($=48.0h$); (c) a magnified view of the region in the panel~(a) marked by the red-dashed square. The colours and the black vectors represent the same as in figure~\ref{fig:instAzy}.}
	\label{fig:instB2}
\end{center}
\end{figure}

Snapshots of instantaneous flow fields obtained in the spanwise-minimal case, Case~B2, are presented in figure~\ref{fig:instB2}. As shown by the middle plane slice ($xz$-plane at $y/h=0.5$) given in the panels~(a) and (c), high- and low-speed streaks are observed with spanwise meanderings. However, there is no very-large-scale structure that fills up the entire channel gap as shown in the panel~(b), unlike the streamwise-minimal and reference cases, clearly due to the narrow spanwise domain size in this case. As the spanwise domain size is half the channel gap $h$, the structures that fit within this spanwise domain width cannot be tall enough to penetrate the whole channel in the wall-normal direction. 

\begin{figure}
\begin{center}
	\includegraphics[width=0.8\hsize]{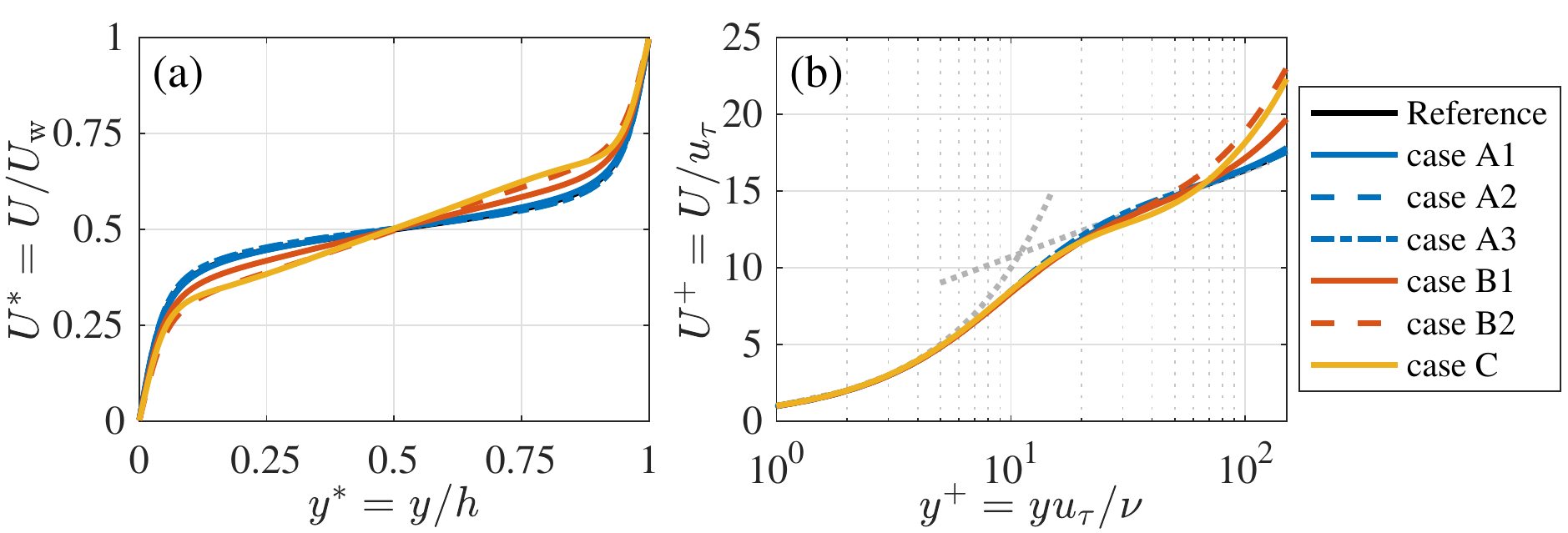}
	\caption{Mean streamwise velocity profiles obtained with different computational domain sizes presented in (a) outer and (b) inner scalings. The colours of the lines represent different series of the computations: blue, series of Case A; red, Case B; yellow, Case C; black, the reference case. The grey dashed lines in the panel (b) represent $U^+=y^+$ and $U^+=1/\kappa \log y^+ +B$ with $\kappa=0.41$ and $B=5.1$. Note that the line for the reference case (black solid line) is hardly visible as that for Case~A1 (blue solid line) is almost on top of it.}
	\label{fig:U}
\end{center}
\end{figure}

Figure~\ref{fig:U} compares the mean streamwise velocity profiles obtained in the various computation runs in (a) the outer and (b) the inner scalings, and one can also find in table~\ref{tab:cond} the values of the friction Reynolds number $Re_\tau=u_\tau (h /2)/\nu$, where $u_\tau$ is the friction velocity defined as $u_\tau = \sqrt{\nu \mathrm{d}U/\mathrm{d}y|_\mathrm{wall}}$. As one can see in figure~\ref{fig:U}, the mean velocity profiles obtained in the series of Case~A are all in a quite good agreement with the reference case, while the results by the series of Case~B and Case~C indicate the influence by reducing $L_z$. The mean velocity profile deviates increasingly from the reference-case results as $L_z$ decreases, clearly corresponding to the disappearance of the very-large-scale structures in these cases as shown in figure~\ref{fig:instB2}. It is also seen in figure~\ref{fig:U}(b) that the logarithmic region of the mean velocity profile is hardly visible in these cases. Consistently with these mean velocity profiles, the $Re_\tau$ values obtained in Cases~A1--3 agree well with the reference value (see table~\ref{tab:cond}); even in the case of the minimal streamwise domain length (Case~A3) the deviation from the reference value is only 0.5\%. On the other hand, $Re_\tau$ obtained in Cases~B1 and B2 clearly decrease as the spanwise domain length $L_z$ decreases.

\begin{figure}
\begin{center}
	\includegraphics[width=0.75\hsize]{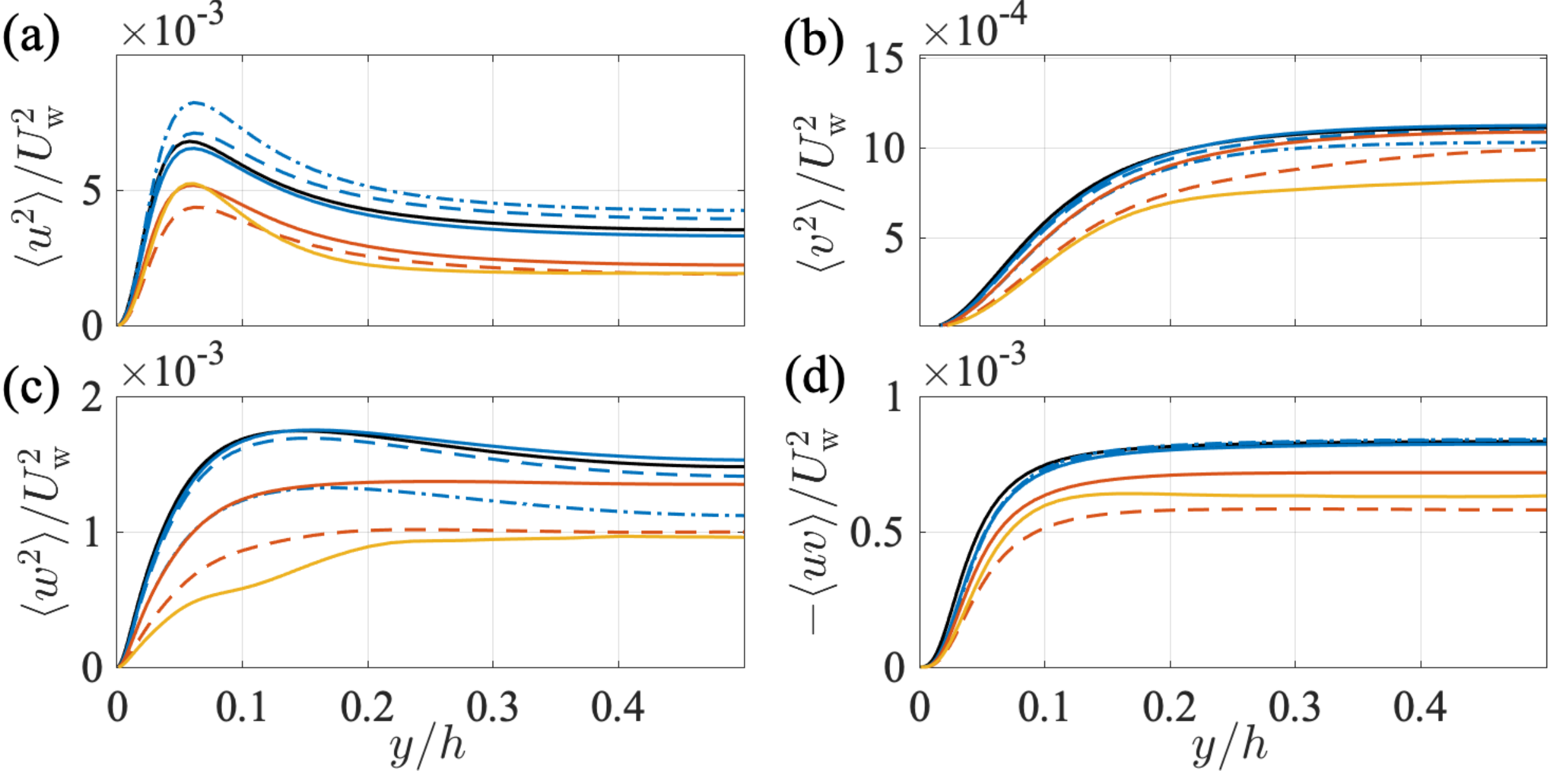}
	\caption{Profiles of the Reynolds stresses scaled by $U_\mathrm{w}^2$ obtained with different computational domain sizes: (a) $\ave{u^2}$, (b) $\ave{v^2}$, (c) $\ave{w^2}$ and (d) $-\ave{uv}$. The colours and styles of the lines represent the same computation cases as in figure~\ref{fig:U}.}
	\label{fig:rs}
\end{center}
\end{figure}

Figure~\ref{fig:rs} presents the profiles of the Reynolds stresses $\ave{u^2}$, $\ave{v^2}$, $\ave{w^2}$ and $-\ave{uv}$ scaled by the wall speed $U_\mathrm{w}$. The Reynolds stresses are clearly affected by reducing $L_x$ as well as $L_z$, unlike the mean velocity profile. As shown in figures~\ref{fig:rs}(a--c) reducing $L_z$ to the minimal size $L_z/h=0.5$ clearly suppresses the fluctuations of all velocity components, corresponding to the suppression of the very-large-scale structures in these cases. On the other hand, reducing $L_x$ increases the streamwise velocity fluctuation $\ave{u^2}$ while decreases the wall-normal and spanwise components $\ave{v^2}$ and $\ave{w^2}$. In particular, the wall-parallel components $\ave{u^2}$ and $\ave{w^2}$ are affected notably, while the influence on the wall-normal component $\ave{v^2}$ is rather moderate. It also should be mentioned here that the sum of these fluctuations, $\ave{u^2} + \ave{v^2} + \ave{w^2}$, is less affected by the change in $L_x$ (not shown), indicating that the energy redistribution between them is suppressed by reducing $L_x$. Such tendencies agree well with the DNS of turbulent channel flow with streamwise-minimal domains by \cite{abe_2018}. The profiles of the Reynolds shear stress $-\ave{uv}$ are also given in figure~\ref{fig:rs}(d). As shown here the profile of $-\ave{uv}$ is insensitive to $L_x$, while reducing $L_z$ remarkably suppresses $-\ave{uv}$, clearly corresponding, again, to the disappearance of the very-large-scale structures. 

\begin{figure}
\begin{center}
	\includegraphics[width=0.95\hsize]{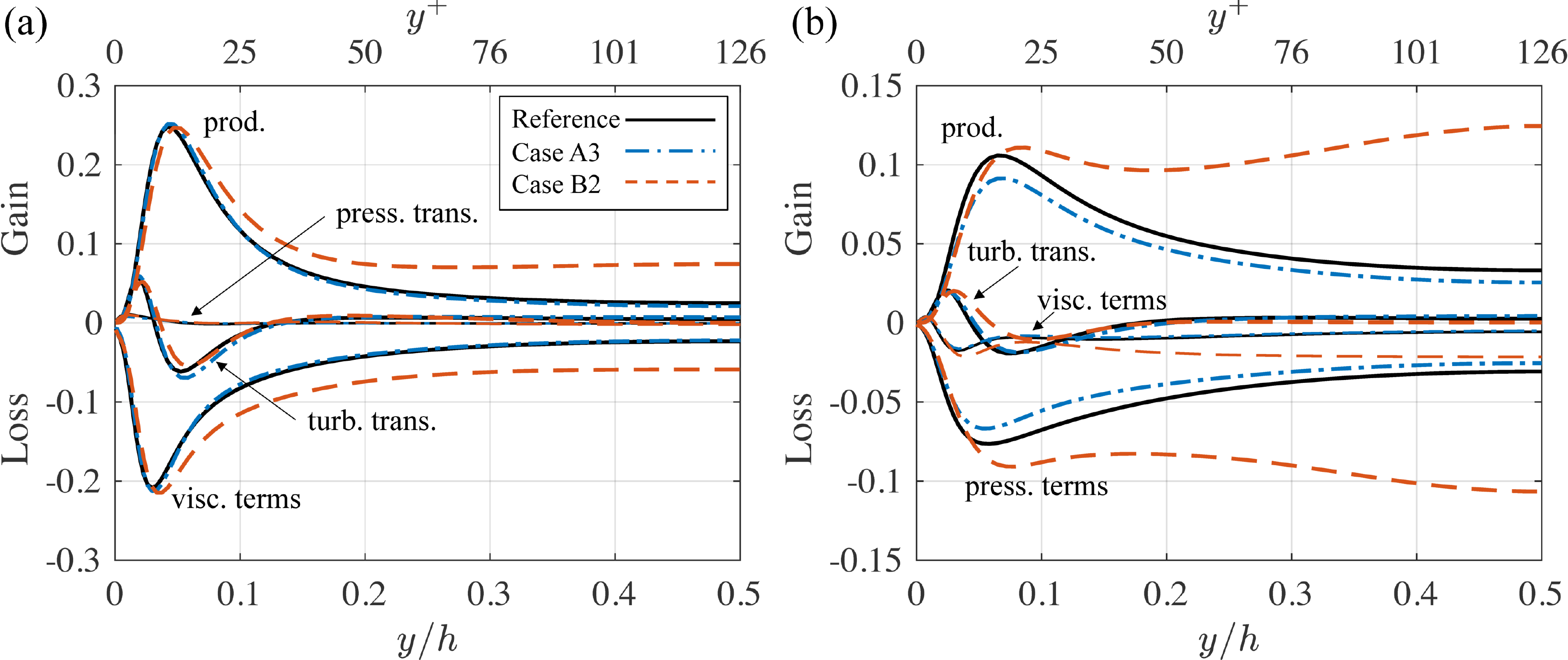}
	\caption{Transport budget of the turbulent kinetic energy $k_\mathrm{t}=\ave{u_i u_i}+\ave{v^2}+\ave{w^2})/2$ and the Reynolds shear stress $-\ave{uv}$ scaled by $u_\tau^4 / \nu$: black solid lines, the reference case; blue chained lines, Case~A3; red dashed lines, Case~B2. The visc.~terms and the press.~terms represents the sum of the viscous dissipation and diffusion terms and that of the pressure transport and pressure-strain correlation terms, respectively. The pressure transport term in (a) and viscous terms in (b) are presented with thiner lines for easier distinction. The upper abscissa of each panel represents the wall-normal height in wall units, $y^+=y u_\tau/\nu$, evaluated based on the results of the reference case.}
	\label{fig:budget} 
\end{center}
\end{figure}

The budget of the transport equation of the turbulent kinetic energy $k_t = (\ave{u^2}+\ave{v^2}+\ave{w^2})/2$ and the Reynolds shear stress $-\ave{uv}$ obtained in the streamwise- and spanwise-minimal cases are given in figure~\ref{fig:budget}. Here, the transport equation of the Reynolds stress $\ave{u_i u_j}$ is defined as
\begin{eqnarray}
\left( \pd{}{t} + U_k \pd{}{x_k} \right) \ave{u_i u_j} =  P_{ij} - \varepsilon_{ij} + \Pi_{ij} + D^\nu_{ij} + D^p_{ij} + D^t_{ij}, \label{eq:rs}
\end{eqnarray}
where the terms on the right-hand side are the production ($P_{ij}$), viscous dissipation ($\varepsilon_{ij}$), pressure-strain correlation ($\Pi_{ij}$), viscous diffusion ($D^\nu_{ij}$), pressure transport ($D^p_{ij}$) and turbulent transport ($D^t_{ij}$) terms, which are defined, respectively, as
\begin{eqnarray}
P_{ij} &=& - \ave{u_i u_k} \pd{U_j}{x_k} - \ave{u_j u_k} \pd{U_i}{x_k}, \quad \varepsilon_{ij} = 2 \nu \ave{\pd{u_i}{x_k} \pd{u_j}{x_k}}, \nonumber \\
\Pi_{ij} &=& \frac{1}{\rho} \ave{p \left( \pd{u_i}{x_j} + \pd{u_j}{x_i} \right)}, \quad D^\nu_{ij} = \nu \pd{^2 \ave{u_i u_j}}{x_k^2}, \nonumber \\
D^p_{ij} &=& - \frac{1}{\rho} \pd{}{x_k} \left( \ave{p u_i} \delta_{ik} + \ave{pu_j} \delta_{jk} \right), \quad
D^t_{ij} = -\pd{\ave{u_i u_j u_k}}{x_k}. \nonumber 
\end{eqnarray}
The profiles of these terms for the transport equation of $k_t$ scaled by the wall units are presented in figure~\ref{fig:budget}(a). Note that the pressure-strain correlation is not presented here as $\Pi_{uu}+\Pi_{vv}+\Pi_{ww}=0$. As shown, any significant difference between the streamwise-minimal and reference cases is not observed for any term, whereas the budget given by the spanwise-minimal case gives clearly larger production and viscous terms in magnitude particularly in the channel central region. The budget of the Reynolds-shear-stress transport is also presented in figure~\ref{fig:budget}(b). In the streamwise-minimal case, while the magnitude of the production $P_{-uv}$ and the pressure-related terms $\Pi_{-uv}+D^p_{-uv}$ are somewhat under estimated as compared to the reference-case results, the other terms are in a quantitative agreement. In the budget given by the spanwise-minimal case, the production and the pressure-related terms are overestimated, similarly to the production and viscous terms in the turbulent kinetic energy budget, and it is also observed that the spatial transport term $D^t_{-uv}$ shows a certain discrepancy from the reference case.  

\begin{figure}
\begin{center}
	\includegraphics[width=0.425\hsize]{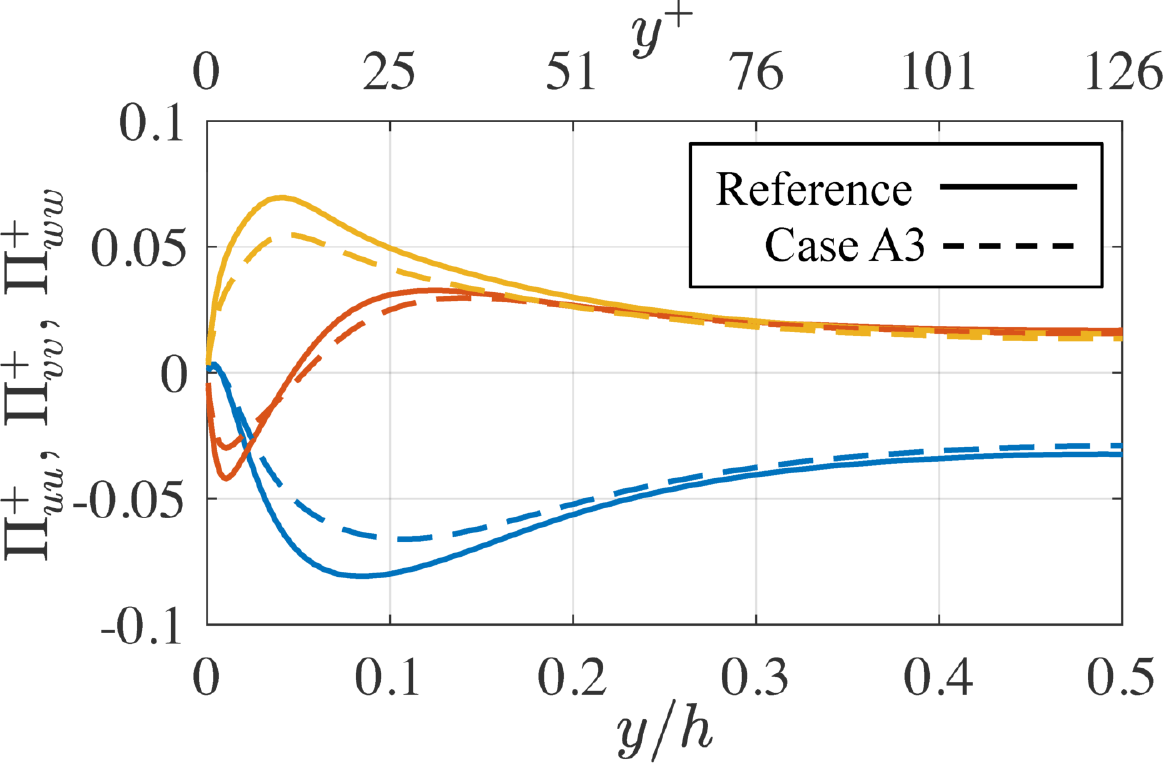}
	\caption{Profiles of the pressure-strain redistribution terms $\Pi_{uu}$ (blue) , $\Pi_{vv}$ (red) and $\Pi_{ww}$ (yellow), comparing the results obtained in the reference case (solid lines) and Case~A3 (dashed lines). The values are scaled by $u_\tau^4/\nu$, and the upper abscissa represents the wall-normal height in wall units, $y^+=y u_\tau/\nu$, evaluated based on the results of the reference case.}
	\label{fig:psij} 
\end{center}
\end{figure}

Figure~\ref{fig:psij} presents the profiles of the pressure-strain correlations $\Pi_{uu}$, $\Pi_{vv}$ and $\Pi_{ww}$, comparing the results of the reference and streamwise-minimal (A3) cases. As shown here the magnitude of all components of the pressure-strain correlations is smaller in Case~A3 than in the reference case. This indicates that the inter-component energy transfer from $\ave{u^2}$ to the other components is suppressed in Case~A3, which is attributable to the observation in figure~\ref{fig:rs} that $\ave{u^2}$ obtained in this case is lager and the other components are smaller than those in the reference case. The same tendency was also reported in the earlier simulations of turbulent channel flow~\citep{abe_2016,abe_2018}. Such suppression of the pressure-strain correlation in the streamwise-minimal case is addressed in detail based on the spectral analysis in Sec.~\ref{sec:lx}. 

\begin{figure}
\begin{center}
	\includegraphics[width=0.75\hsize]{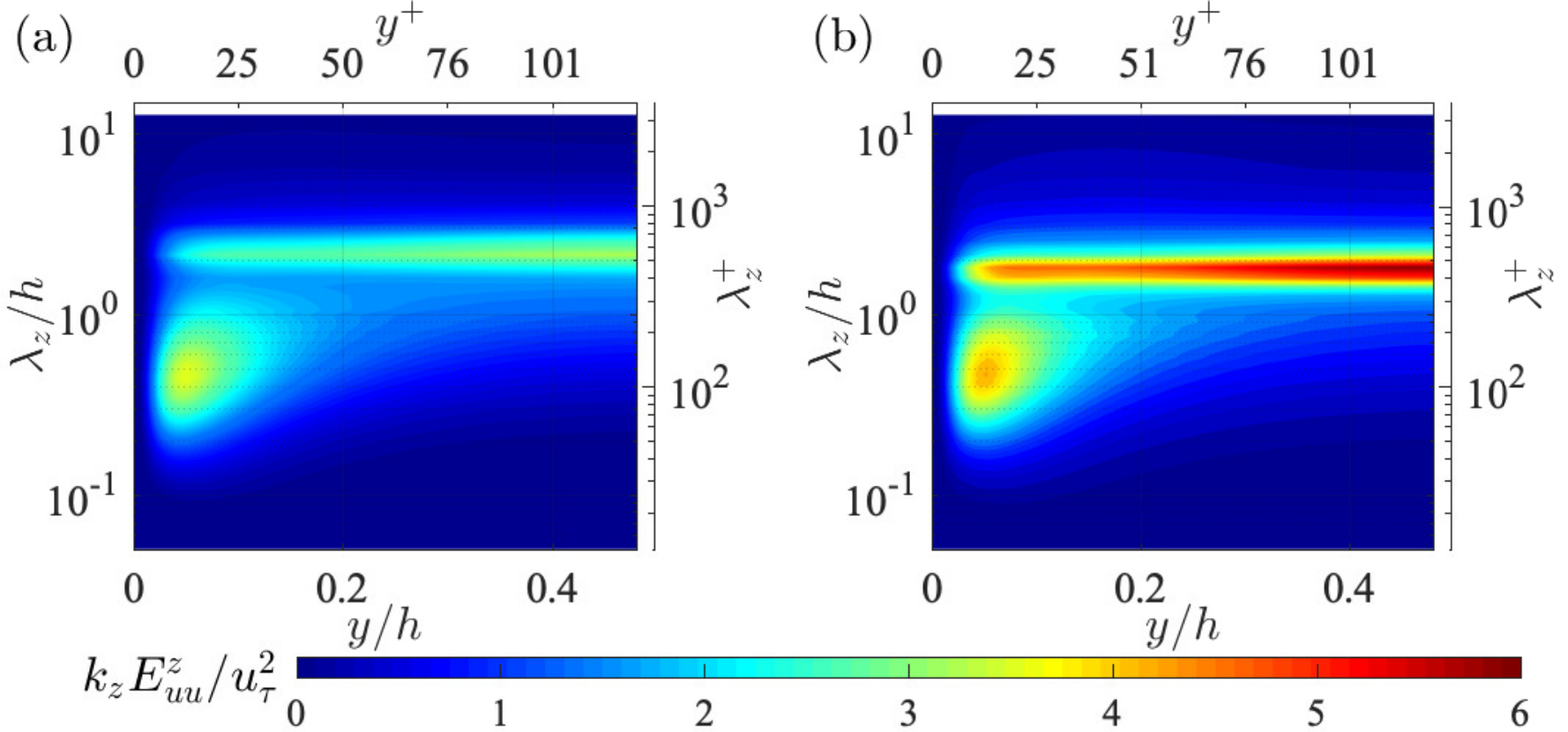}
	\caption{Space-wavelength ($y$-$\lambda_z$) diagrams of the spanwise one-dimensional spectrum of the streamwise turbulent energy $E_{uu}^z$; (a) the reference case; (b) Case A3 (the streamwise-minimal case). The values are scaled by $u_\tau^2$.}
	\label{fig:ezuu} 
\end{center}
\end{figure}

The streamwise-minimal (A3) case successfully reproduces the reference-case results also in terms of the spectral energy distributions. Figure~\ref{fig:ezuu} presents the distributions of the premultiplied spanwise one-dimensional spectra of the streamwise turbulent energy $k_z E^z_{uu}(y,k_x)$, comparing the results obtained in the reference and the streamwise-minimal cases. Here $k_z$ is the spanwise wavenumber $k_z=2\pi/\lambda_z$, where $\lambda_z$ is the spanwise wavelength. As shown in the figure, the spectrum distributions obtained in these two cases are in a fairly good agreement in that both indicate two energy peaks: one located at near-wall region at relatively small wavelengths $\lambda_z^+=\lambda_z u_\tau /\nu \approx 100$ and the other located at the channel centre at large wavelengths $\lambda_z/h \approx 2$. In particular, the energy peak at the larger wavelength spread broadly in $y$-direction and reaches the near-wall region where the near-wall energy peak is located at smaller wavelength. These spectral energy peaks clearly corresponds to the near-wall and the very-large-scale structures observed in figure~\ref{fig:instAzy}(a). As described here, in turbulent plane Couette flow the energy peaks representing the inner and outer structures are observed clearly separated even at the relatively low Reynolds number $Re_\tau \approx 126$ investigated in the present study, which is due to the non-zero mean velocity gradient at the channel centre.

As described so far, the streamwise-minimal case reproduces the reference-case results fairly well despite the substantially limited degree of freedom in the streamwise direction, while the effect of reducing $L_z$ is shown to be, as expected, remarkable mainly because the very-large-scale structure disappears with insufficient spanwise domain width. In the following section, the results provided by the streamwise- and spanwise-minimal domains are further examined based on the spectral analysis of the Reynolds stress transport. The focus is put, in the streamwise-minimal case, on why the streamwise-minimal domain still successfully reproduces the reference case results, while regarding the spanwise-minimal case the role of interaction between the near-wall and very-large-scale structures in the Reynolds-stress transport is addressed. 

\subsection{Spectral analysis on the effect of reducing $L_x$}
\label{sec:lx}

\begin{figure}
\begin{center}
	\includegraphics[width=0.9\hsize]{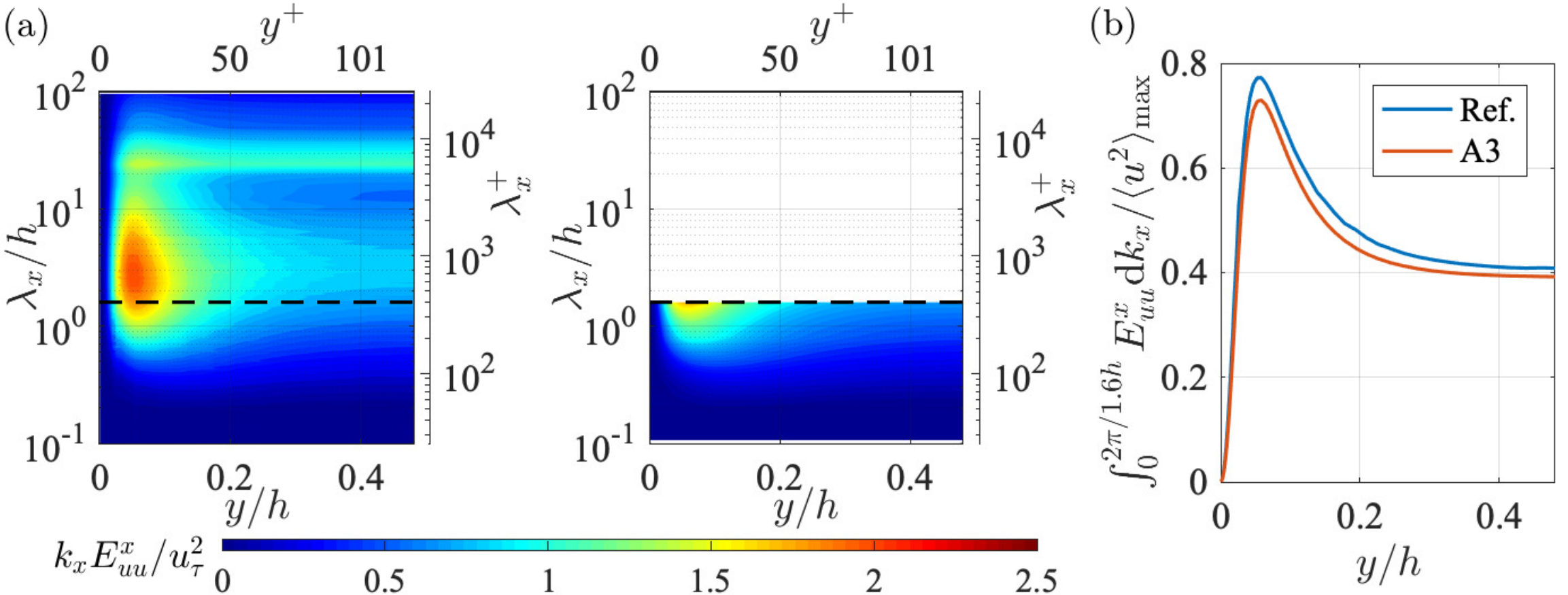}
	\caption{(a) Space-wavelength ($y$-$\lambda_x$) diagrams of premultiplied streamwise one-dimensional spectrum of the streamwise turbulent energy $k_x E^x_{uu}$ obtained in (left) the reference and (right) the streamwise-minimal (A3) cases. The horizontal black dashed lines represent the streamwise domain size in the streamwise-minimal case $\lambda_x/h=1.6$. (b) Profiles of streamwise turbulent energy spectra integrated for the wavelength range $\lambda_x/h > 1.6$ (i.e.~$0 < k_xh < 2\pi/1.6$); (blue) the reference case, (red) the streamwise-minimal (A3) case. Note that, in Case~A3, $\int^{2\pi/1.6h}_0 E^x_{uu}(y,k_x) \mathrm{d} k_x = E^x_{uu}(y,0) \Delta k_x$, where $\Delta k_x = 2 \pi/L_x$, since the $x$-independent mode is the only Fourier mode in the integrated $k_x$ range due to the limited domain size $L_x$. $\ave{u^2}_\mathrm{max}$ is the maximum value of the $\ave{u^2}$ profile presented in figure~\ref{fig:rs}(a).}
	\label{fig:exuu}
\end{center}
\end{figure}

In this section, we analyse the streamwise spectra of the Reynolds stresses and their transport in the streamwise-minimal case. Figure~\ref{fig:exuu}(a) compares the space-wavelength ($y$-$\lambda_x$) diagrams of the premultiplied streamwise one-dimensional spectrum of the streamwise turbulent energy $k_x E^x_{uu}(y,k_x)$. As shown here the $E^x_{uu}$ distribution obtained in the reference case shows both energy peaks corresponding to the near-wall structure (the one located at small wavelength $\lambda_x^+  \approx 600$) and the very-large-scale structure (the energy broad band at large wavelength $\lambda_x/h \approx 48$), similarly to the spanwise energy spectrum $E^z_{uu}(y,k_z)$ in figure~\ref{fig:ezuu}. It should be noted here that not only the energy broad band at large $\lambda_x$ but also the near-wall peak is located in the wavelength range $\lambda_x^+>400$, and therefore in the streamwise-minimal case (A3) both of them are located outside the $y$-$\lambda_x$ diagram of $k_x E^x_{uu}$ presented here. In stead, in the streamwise-minimal case, both the inner and outer peaks of the turbulent energy spectrum are accounted for by the $x$-independent mode, i.e.~the Fourier mode at $k_x=0$ (or at $\lambda_x=\infty$). This is depicted in figure~\ref{fig:exuu}(b), where $E^x_{uu}(y,0) \Delta k_x$ (here $\Delta k_x = 2\pi/L_x$), i.e.~the amount of the streamwise turbulent energy on the Fourier mode at $k_x=0$, of the streamwise-minimal case is compared to the energy integrated over the range $\lambda_x/h > 1.6$:
\begin{equation}
 \int^\infty_{\log 1.6h} k_x E^x_{uu} \mathrm{d} (\log\lambda_x) \left(= \int^{2\pi/1.6h}_0 E^x_{uu} \mathrm{d}k_x \right)\label{eq:ex}
\end{equation}
 in the reference case. As shown, the $x$-independent Fourier mode in the streamwise-minimal case accounts for the equivalent fraction of the total $\ave{u^2}$, up to 80~\% of $\ave{u^2}_\mathrm{max}$, to the $\lambda_x/h>1.6$ range in the reference case. This means that the high- and low-$u$ streaks in both the near-wall and channel-core regions are $x$-independent in the streamwise-minimal case. 

\begin{figure}
\begin{center}
	\includegraphics[width=1\hsize]{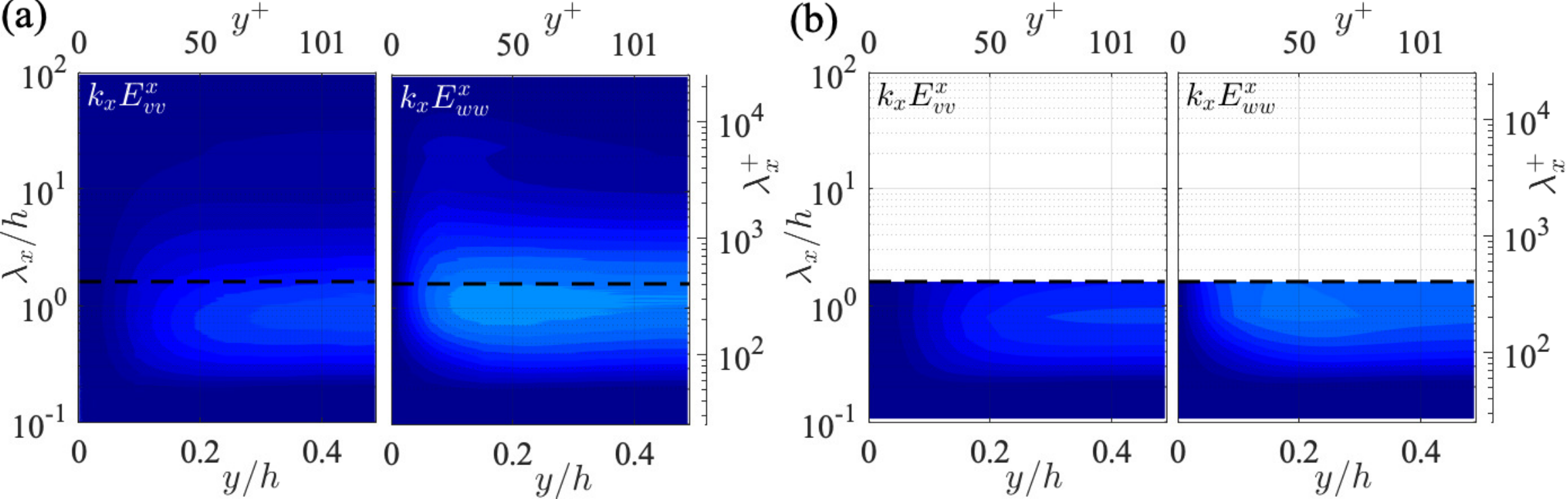}
	\caption{Space-wavelength ($y$-$\lambda_x$) diagrams of premultiplied streamwise one-dimensional spectrum of the wall-normal turbulent energy $k_x E^x_{vv}$ and the spanwise turbulent energy $k_z E^x_{ww}$ obtained in (a) the reference case and (b) the streamwise-minimal (A3) case. The values are scaled by $u_\tau^2$, and the colour scale is the same as in figure~\ref{fig:exuu}(a).}
	\label{fig:exvv}
\end{center}
\end{figure}

The streamwise one-dimensional spectra of the cross-streamwise velocity fluctuations $E^x_{vv}$ and $E^x_{ww}$ obtained in the reference and streamwise-minimal cases are also presented in figure~\ref{fig:exvv}. Note that the colour scale for the $k_x E^x_{vv}$ and $k_x E^x_{ww}$ distributions shown here is the same as in figure~\ref{fig:exuu}(a), showing that the magnitudes of $E^x_{vv}$ and $E^x_{ww}$ are relatively small as compared to the streamwise turbulent energy $E^x_{uu}$. The results obtained in the reference case given in figure~\ref{fig:exvv}(a) show that the spectral energies of $\ave{v^2}$ and $\ave{w^2}$ are mainly distributed in the relatively small wavelength range $\lambda_x/h<1.6$, and therefore their distributions are not significantly affected in the streamwise-minimal case as shown in figure~\ref{fig:exvv}(b), unlike the distribution of $E^x_{uu}$. 

The transport equation of the Reynolds stress spectra can be derived by decomposing the Reynolds-stress transport equation (\ref{eq:rs}) into spectral contribution from each wavenumber as done by, for example, \cite{mizuno_2016}, \cite{lee_2015b,lee_2019} and \cite{kawata_2018,kawata_2019}. The transport equation of the streamwise one-dimensional spectra of the Reynolds stress $E^x_{ij}$ is written as
\begin{eqnarray}
\left( \pd{}{t} + U_k \pd{}{x_k} \right) E^x_{ij} = pr^x_{ij} - \epsilon^x_{ij} + \pi^x_{ij} + d^{p,x}_{ij} + d^{\nu,x}_{ij} + d^{t,x}_{ij} + tr^x_{ij} \label{eq:exij},
\end{eqnarray}
where the first six terms on the right-hand side represent the spectral contribution from each streamwise wavenumber to the corresponding terms in the overall transport equation~(\ref{eq:rs}), whereas the last term $tr^x_{ij}$ is an additional term that represents the interscale transfer of the Reynolds stress between different streamwise wavenumbers. Here the spectral Reynolds-stress transport equation~(\ref{eq:exij}) is derived by the same procedure as done by \cite{kawata_2019}, which is briefly described in Appendix~\ref{app:a}.

\begin{figure}
\begin{center}
	\includegraphics[width=0.9\hsize]{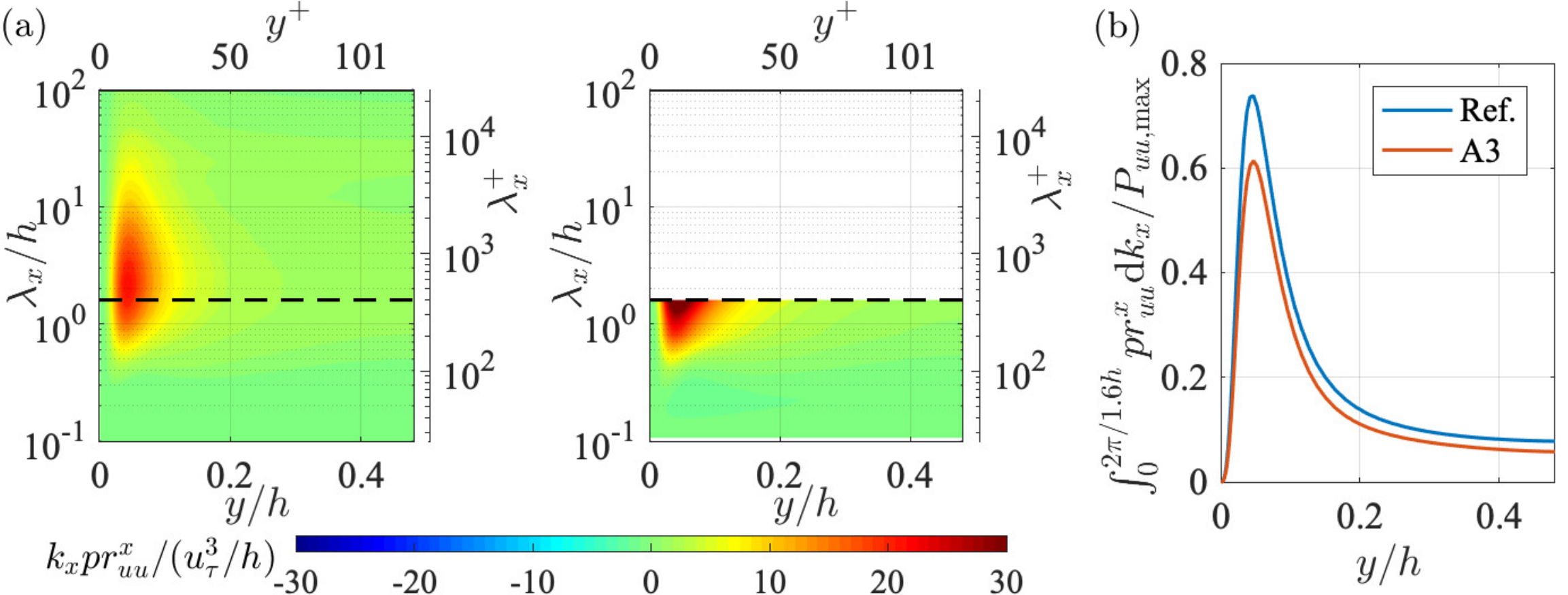}
	\caption{Distributions of streamwise one-dimensional spectrum of the streamwise turbulent energy production $pr^x_{uu}$ obtained in the reference and streamwise-minimal (A3) cases, presented in the same manner as in figure~\ref{fig:exuu}: (a) $y$-$\lambda_x$ diagrams of $k_x pr^x_{uu}$; (b)~profiles of $pr^x_{uu}$ integrated for $\lambda_x/h > 1.6$ ($0<k_xh<2\pi/1.6$). Note that, in Case~A3, $\int^{2\pi/1.6h}_0 pr^x_{uu}(y,k_x) \mathrm{d} k_x=pr^x_{uu}(y,0) \Delta k_x$. $P_{uu,\mathrm{max}}$ represents the maximum value of the turbulent energy production~$P_{uu}$.}
	\label{fig:prxuu} 
\end{center}
\end{figure}

Figure~\ref{fig:prxuu}(a) presents the $y$-$\lambda_x$ diagrams of the premultiplied spectral energy production $k_x pr^x_{uu}$ comparing the results by the reference and the streamwise-minimal cases. Similarly to the distribution of the streamwise energy spectrum $E^x_{uu}$, most of the spectral energy production $pr^x_{uu}$ is distributed in the relatively large wavelength range $\lambda_x^+>400$ in the reference case. In the streamwise-minimal case, on the other hand, the corresponding energy productions occur on the $x$-independent mode as shown by figure~\ref{fig:prxuu}(b), where $pr^x_{uu}(y,0) \Delta k_x$ of the streamwise-minimal case is shown to be comparable to the the integrated production $\int^{2\pi/1.6h}_0 pr^x_{uu} \mathrm{d}k_x$ of the reference case. This indicates that the overall production $P_{uu}$ mainly represents the generation of $x$-independent streaks in the streamwise-minimal case.  

\begin{figure}
\begin{center}
	\includegraphics[width=0.9\hsize]{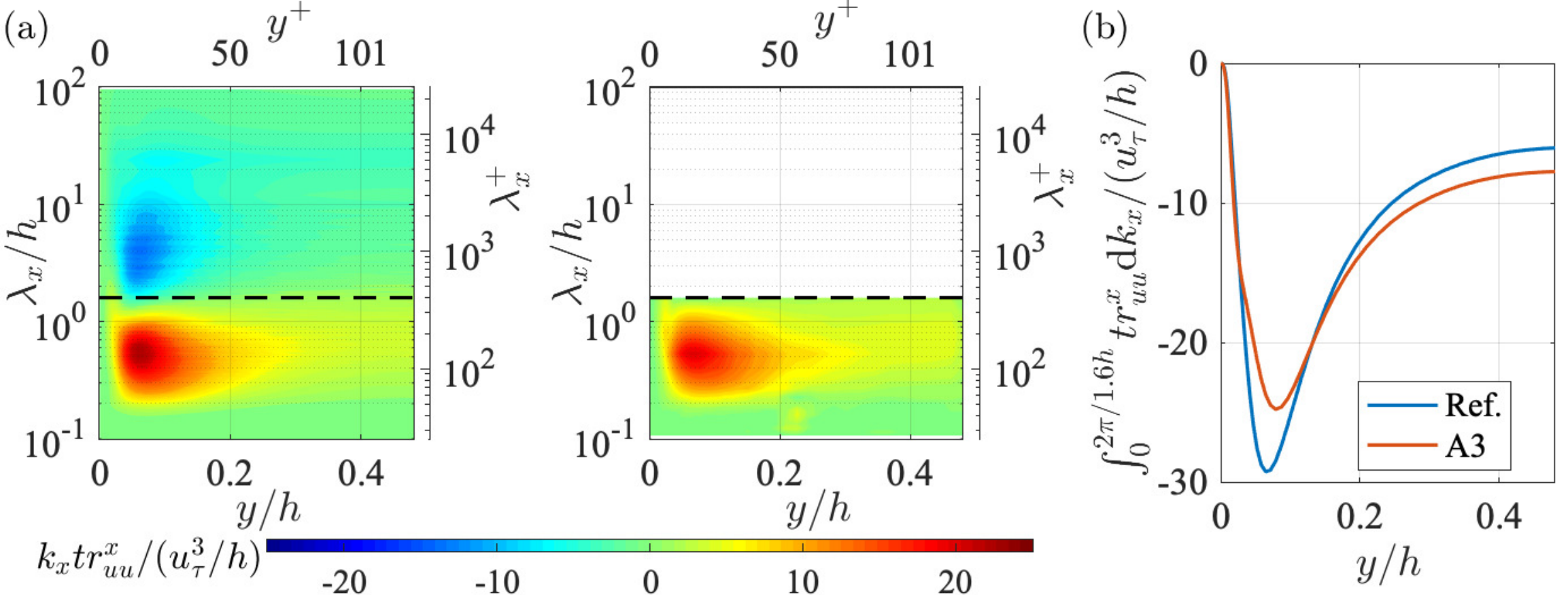}
	\caption{Distributions of the interscale transport of the streamwise turbulent energy in the streamwise wavenumber direction $tr^x_{uu}$ obtained in the reference and streamwise-minimal (A3) cases, presented in the same manner as in figure~\ref{fig:exuu}. Note that, in Case~A3, $\int^{2\pi/1.6h}_0 tr^x_{uu}(y,k_x) \mathrm{d} k_x=tr^x_{uu}(y,0) \Delta k_x$.}
	\label{fig:trxuu}
\end{center}
\end{figure}

Figure~\ref{fig:trxuu} presents  the interscale transfer of the streamwise turbulent energy $tr^x_{uu}$. As shown by the space-wavelength diagrams in the panel~(a), in the reference case the turbulent energy, which is produced by $pr^x_{uu}$ at large wavelengths as presented in figure~\ref{fig:prxuu}(a), is transferred towards smaller wavelengths throughout the channel, and the streamwise-minimal-case result reproduces quite well the $tr^x_{uu}$ distribution of the reference case in the range $\lambda_x/h < 1.6$. It is particularly interesting to note  here that the boundary between the energy-donating and -receiving $\lambda_x$ ranges in the reference case appears to be rather independent of $y$ and coincide with the streamwise minimal domain size $L_x=1.6h$, as indicated by the black dashed line. In the streamwise-minimal case, the energy gained in the $\lambda_x/h<1.6$ range is thoroughly supplied from the $x$-independent mode since this is the only mode in the range $\lambda_x/h > 1.6$. Figure~\ref{fig:trxuu}(b) compares the total amount of energy removed from the wavelength range $\lambda_x/h>1.6$, i.e.~$\int^{2\pi/1.6h}_0 tr^x_{uu} \mathrm{d}k_x$, in the reference case to the energy transfer from the $x$-independent mode $tr^x_{uu}(y,0) \Delta k_x$ in the streamwise-minimal case. As shown here, $tr^x_{uu}(y,0) \Delta k_x$ in the streamwise-minimal case accounts for the equivalent energy transfer to $\int^{2\pi/1.6h}_0 tr^x_{uu} \mathrm{d}k_x$ of the reference case.

The good agreement between the $tr^x_{uu}$ distributions obtained in the reference and streamwise-minimal cases given in figure~\ref{fig:trxuu} may give us an insight into the mechanism of the interscale energy transfer. From the view point of the Fourier mode analysis, the interscale energy transfer is generally caused by triad interactions between three wavenumbers. In the streamwise-minimal case, the $x$-independent mode is the only mode in the range $\lambda_x/h>1.6$ and therefore the energy transfer from the $x$-independent mode to smaller scales across $\lambda_x/h=1.6$ is caused by interaction between $k_x=0$ and non-zero wavenumbers $k_x=\pm 2\pi m/1.6h$ ($m$ is a positive integer). The good agreement between the $tr^x_{uu}$ distribution given in figure~\ref{fig:trxuu}(b) indicates that the energy transfer from larger to smaller scales across $\lambda_x/h=1.6$ in the reference case is also mainly cased by similar direct interactions between far-separated wavenumbers, i.e.~the triad interactions between very small $k_x$ ($\approx0$) and large ($k_x=\pm 2 \pi m/L_x$) wavenumbers; if the other triad combinations of moderate wavenumbers play important roles, such triad interactions cannot be simulated in the streamwise-minimal case and thus the good agreement with the reference case shown in figure~\ref{fig:trxuu}(b) would not have been achieved. 

\begin{figure}
\begin{center}
	\includegraphics[width=0.9\hsize]{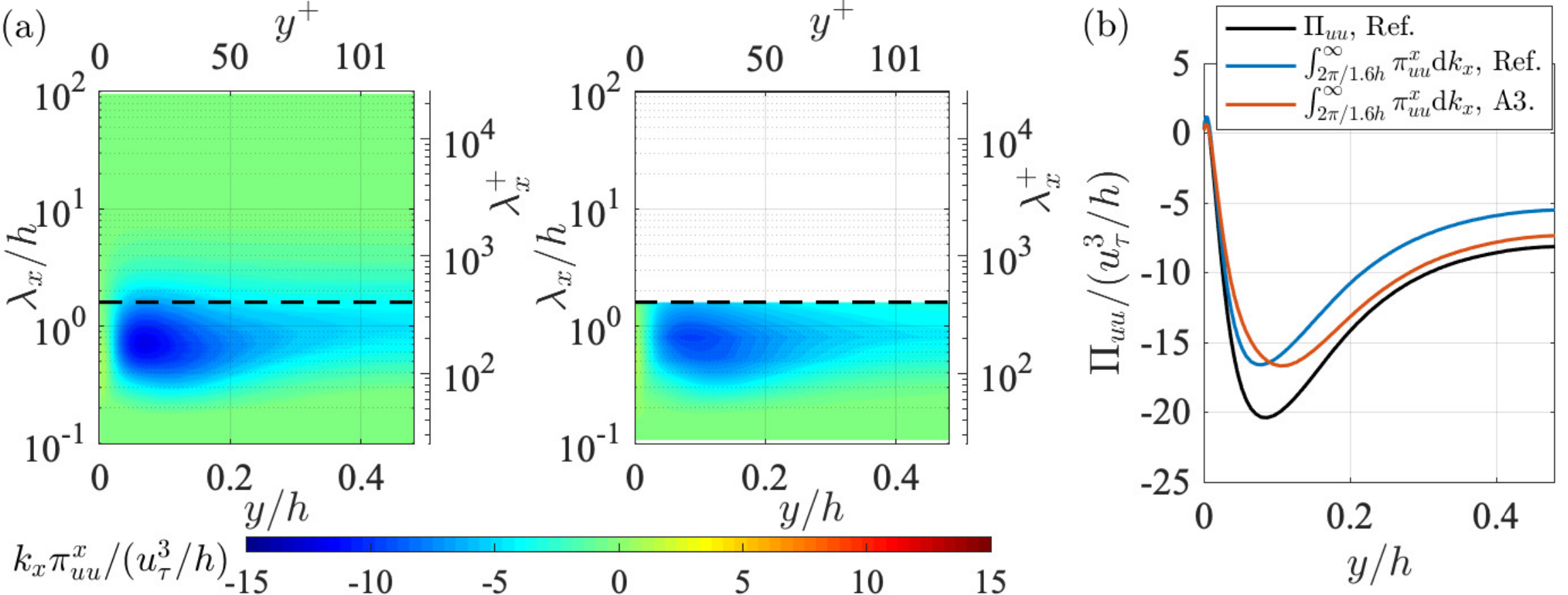}
	\caption{Distributions of streamwise one-dimensional cospectrum of the pressure-strain correlation $\pi^x_{uu}$ obtained in the reference and streamwise-minimal (A3) cases, presented in a similar manner to figure~\ref{fig:exuu}. Note that, in Case~A3, $\int^\infty_{2\pi/1.6h} \pi^x_{uu} (y,k_x) \mathrm{d}k_x = \Pi_{uu}(y)$. The overall pressure-strain correlation $\Pi_{uu}$ by the reference case (black) is also compared in the panel~(b).}
	\label{fig:pixuu}
\end{center}
\end{figure}

Figure~\ref{fig:pixuu} presents the distribution of the pressure-strain cospectrum $\pi^x_{uu}$ comparing the results obtained in the reference and the streamwise-minimal cases.  One can see here that the pressure-strain cospectrum is mostly significant in the relatively small wavelength range $\lambda_x/h < 1.6$ throughout the channel, and the $\pi^x_{uu}$ distribution obtained in the streamwise-minimal case reproduces the result of the reference case fairly well. This tendency of $\pi^x_{uu}$ is consistent with the distributions of $E^x_{vv}$ and $E^x_{ww}$ shown in figure~\ref{fig:exvv}. It also should be noted here that the $\pi^x_{uu}$ distribution of the reference case shows a certain contribution from the range $\lambda_x/h >1.6$, which cannot be taken into account in the streamwise-minimal case because $\pi^x_{uu}=0$ at $k_x=0$ as $\partial u/\partial x = 0$ for any $x$-independent structure. The contribution from the relatively small $\lambda_x$ range $\lambda_x/h<1.6$ is evaluated as $\int^{\infty}_{2\pi/1.6h} \pi^x_{uu} \mathrm{d} k_x$ for both the reference and streamwise-minimal cases and compared in figure~\ref{fig:pixuu}(b). Note here that for the streamwise-minimal case $\int^{\infty}_{2\pi/1.6h} \pi^x_{uu} \mathrm{d} k_x=\int^{\infty}_{0} \pi^x_{uu} \mathrm{d} k_x=\Pi_{uu}$ since the contribution from the rage $\lambda_x/h>1.6$ ($0<k_xh<1/1.6$) is zero as explained above. As shown, the profile of $\int^{\infty}_{2\pi/1.6h} \pi^x_{uu} \mathrm{d} k_x$ given by the reference case is, of course, somewhat smaller in magnitude than the overall pressure-strain correlation $\Pi_{uu}$ of the reference case, and $\Pi_{uu}$ obtained in the streamwise-minimal case shows the equivalent magnitude to $\int^{\infty}_{2\pi/1.6h} \pi^x_{uu} \mathrm{d} k_x$ of the reference case. This indicates that the suppression of $\Pi_{uu}$ in the streamwise-minimal case observed in figure~\ref{fig:psij} is mainly because the energy-containing streaks are forced to $x$-independent due to the limited streamwise domain size, and therefore do not induce the energy redistribution from $\ave{u^2}$ to the other components by itself. 

The spectral analysis on the turbulent energy transport described above has shown that, in the reference case, the streamwise turbulent energy $\ave{u^2}$ is produced mainly at relatively large streamwise wavelengths $\lambda_x^+>400$, and then transferred by the interscale transport $tr^x_{uu}$ towards smaller wavelength range where the energy is redistributed to the cross-streamwise components by the pressure-strain correlation. In the streamwise-minimal case, the energy production mainly occurs at $k_x=0$ in both the near-wall and core regions of the channel, indicating that both streaks of the near-wall and very-large-scale structures are forced to be $x$-independent, and the energy is transferred from the $x$-independent mode to all other modes in the range $\lambda_x^+<400$. Importantly, in spite of such effect of limited degree of freedom the streamwise-minimal domain still reproduces the flow field obtained in the reference case fairly well as shown in Sec.~\ref{sec:flow}. This indicates that the interactions between the $x$-independent streaks and relatively small structures in the range $\lambda_x^+<400$ are the essential dynamics for both the inner and outer structures, and the spanwise meanderings of their streaks observed in the reference case (such as those presented in figure~\ref{fig:instAzy}(b)) are, therefore, rather not important feature of wall turbulence, as previously suggested by \cite{abe_2018}. Furthermore, as discussed above the good agreement between the $tr^x_{uu}$ distributions shown in figure~\ref{fig:trxuu}(b) indicates that  $tr^x_{uu}$ mainly represents the direct energy transfers between far separated scales, i.e.~very large ($\lambda_x \gg h$) and relatively small ($\lambda_x^+<400$) wavelengths via, for instance instabilities, rather than the successive energy transfers from one scale to neighbouring a-little-smaller one by the turbulent energy cascade. 

\subsection{Spectral analysis on the effect of reducing $L_z$}
\label{sec:lz}

Next, we investigate the Reynolds stress transports in the spanwise-minimal (B2) case with emphasis on the difference in the interscale transfer effects between the reference and spanwise-minimal cases, in order to elucidate the role of the interaction between the inner and outer structures. In this investigation we focus the spanwise spectra of the Reynolds stresses $E^z_{ij}(y,k_z)$. Their transport equation is written as
\begin{eqnarray}
\left( \pd{}{t} + U_k \pd{}{x_k} \right) E^z_{ij} = pr^z_{ij} - \epsilon^z_{ij} + \pi^z_{ij} + d^{p,z}_{ij} + d^{\nu,z}_{ij} + d^{t,z}_{ij} + tr^z_{ij} \label{eq:ezij},
\end{eqnarray}
similarly to Eq.~(\ref{eq:exij}). The first six terms on the right-hand size now represent the spectral contribution from each spanwise wavenumber to the corresponding term in (\ref{eq:rs}), and the last term $tr^z_{ij}$ indicates the interscale transfer in the spanwise wavenumber direction. 

\begin{figure}
\begin{center}
	\includegraphics[width=0.9\hsize]{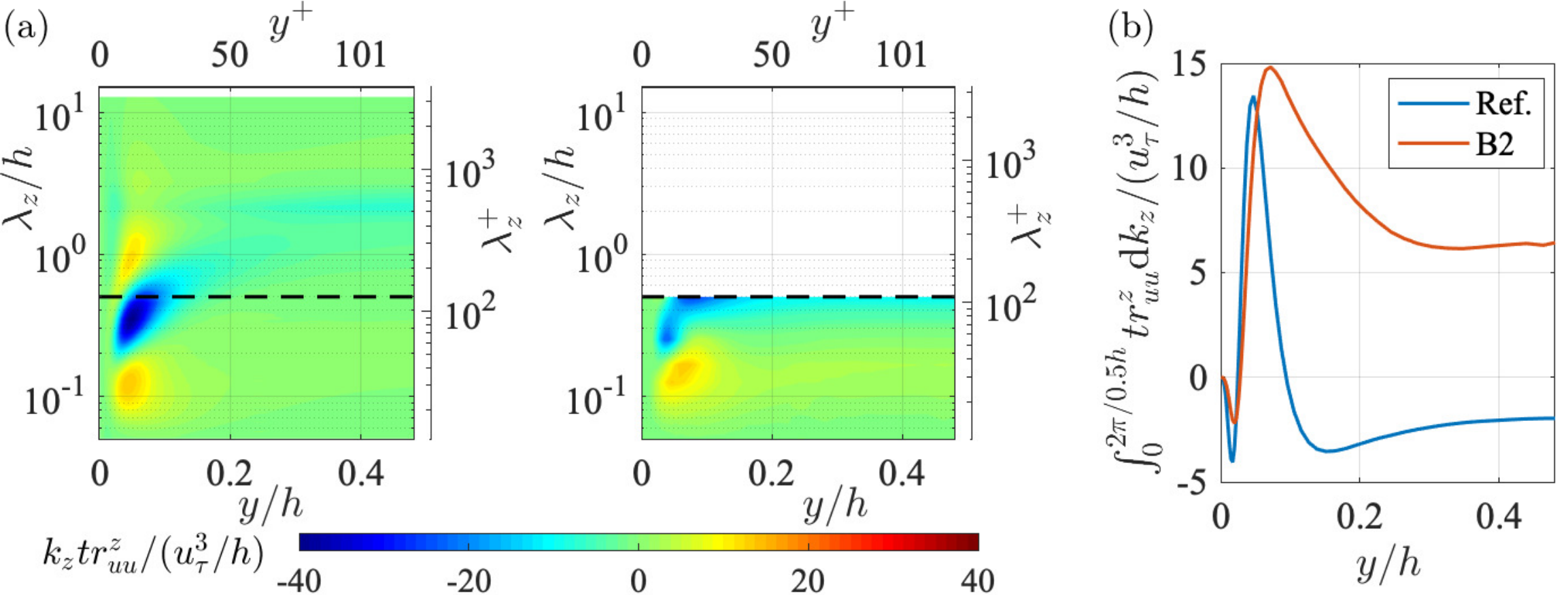}
	\caption{(a) Space-wavelength ($y$-$\lambda_z$) diagrams of the premultiplied interscale transport of the streamwise turbulent energy in the spanwise wavenumber direction $k_z tr^z_{uu}$ obtained in (left) the reference and (right) the spanwise-minimal (B2) cases. The horizontal black dashed lines represent the spanwise domain size in the spanwise-minimal case $\lambda_z/h=0.5$. (b) Profiles of the interscale transport $tr^z_{uu}$ integrated for the wavelength range $\lambda_z/h > 0.5$ (i.e.~$0 < k_zh < 2\pi/0.5$); (blue) the reference case, (red) the spanwise-minimal (B2) case. Note that ,in Case~B2, $\int^{2\pi/0.5h}_0 tr^z_{uu}(y,k_z) \mathrm{d} k_z=tr^z_{uu}(y,0) \Delta k_z$, where $\Delta k_z=2 \pi/L_z$, due to the limited domain size~$L_z$.}
	\label{fig:trzuu}
\end{center}
\end{figure}

Figure~\ref{fig:trzuu}(a) presents the $y$-$\lambda_z$ diagrams of the interscale transport of the streamwise turbulent energy $tr^z_{uu}$. As one can see here, the $tr^z_{uu}$ distribution obtained in the reference case indicates that $\ave{u^2}$ is transferred mainly from larger ($\lambda_z/h\approx2$) to smaller $\lambda_z$, while in the near-wall region the transport from smaller to larger $\lambda_z$ is also found. This is in contrast to the interscale transfer in the streamwise wavenumber direction $tr^x_{uu}$, which indicates only the transport from larger to smaller wavelength $\lambda_x$ throughout the channel as presented in figure~\ref{fig:trxuu}(a). This point is further addressed later in Sec.~\ref{sec:trz}. In the spanwise-minimal case (Case~B2), the $tr^z_{uu}$ distribution reproduces the reference-case result fairly well for the wavelength range $\lambda_z/h<0.5$. The energy is mainly removed from $\lambda_z/h=0.5$ and partly transferred to smaller $\lambda_z$ in the near-wall region, and the rest is transferred to the $z$-independent mode, as indicated by the large positive values of $tr^z_{uu}(y,0) \Delta k_z$ of the spanwise-minimal case presented in figure~\ref{fig:trzuu}(b). The $tr^z_{uu}(y,0) \Delta k_z$ profile is compared with the energy amount transferred from the range $\lambda_z/h<0.5$ to $\lambda_z/h>0.5$ in the reference case, and one can see that the reversed energy transfer to the $z$-independent mode in the spanwise-minimal case captures qualitatively the tendency of the corresponding interscale energy transfer in the reference case, particularly the inverse interscale transfer in the near-wall region. Such an agreement between the reference and spanwise-minimal cases are somewhat surprising, given that the instantaneous flow field obtained in the spanwise-minimal case is qualitatively different from the reference case in that the very-large-scale structure does not exist in the channel core region, as observed in Sec.~\ref{sec:flow}. 

\begin{figure}
\begin{center}
	\includegraphics[width=0.9\hsize]{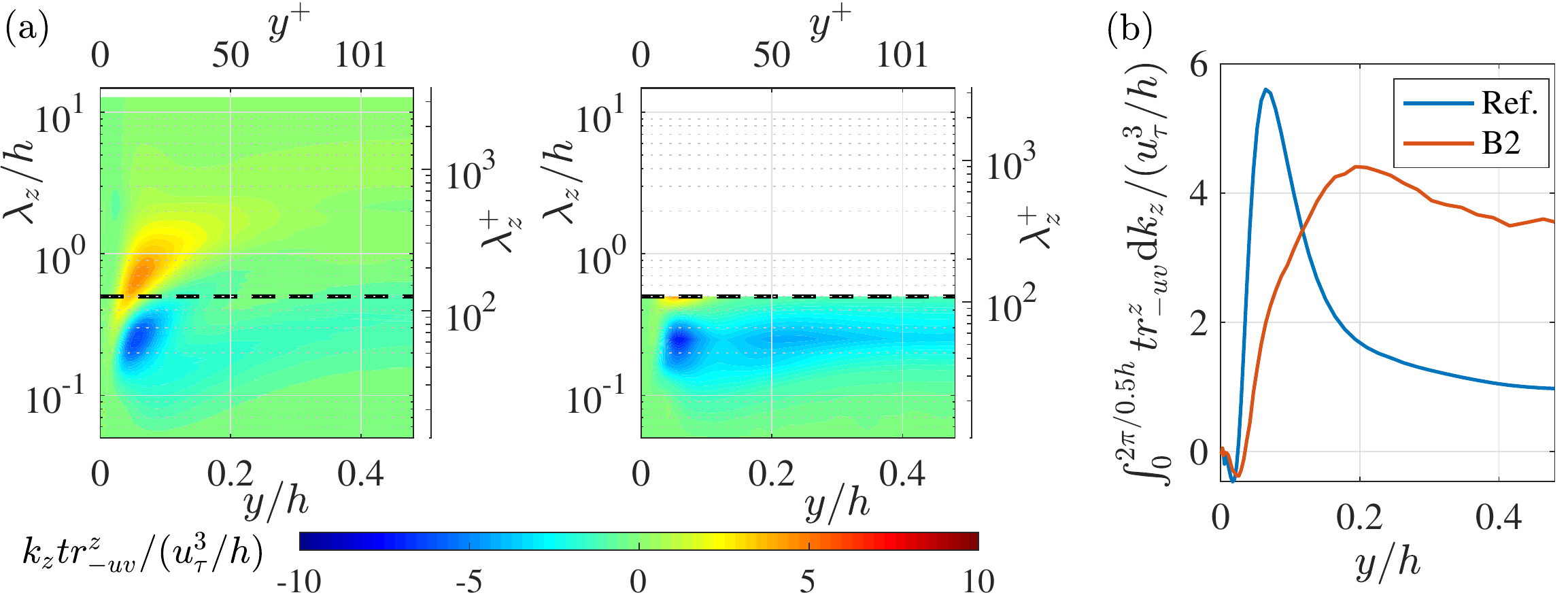}
	\caption{Distributions of the interscale transport of the Reynolds shear stress in the spanwise wavenumber direction $tr^z_{-uv}$ obtained in the reference and spanwise-minimal (B2) cases, presented in the same manner as in figure~\ref{fig:trzuu}. }
	\label{fig:trzuv}
\end{center}
\end{figure}

\cite{kawata_2018} experimentally investigated the spectral transport of the Reynolds stresses in a turbulent plane Couette flow based on the spanwise-Fourier-mode analysis and showed that the Reynolds shear stress $-\ave{uv}$ is transferred from smaller to larger $\lambda_z$ throughout the channel. This inverse interscale transport of the Reynold shear stress is also observed in the present study. As shown in figure~\ref{fig:trzuv}(a), the distributions of the interscale Reynolds-shear-stress transport $tr^z_{-uv}$ obtained in the reference case presents the transfer from smaller to larger $\lambda_z$ throughout the channel, consistently with the experimental observation by \cite{kawata_2018}. The result by the spanwise-minimal case also reproduces the same tendency, where the Reynold shear stress is mainly transferred from around $\lambda_z/h \approx 0.25$ to larger wavelengths. As indicated by the comparison between $\int^{2\pi/0.5h}_0 tr^z_{-uv} \mathrm{d}k_z$ of the reference case and $tr^z_{-uv}(y,0) \Delta k_z$ of the spanwise-minimal case, the interscale transfer between the $\lambda_z/h <0.5$ range and the $z$-independent mode in the spanwise-minimal case represents qualitatively well the inverse interscale transfers of the Reynolds shear stress observed in the reference case.

As described above, the spanwise-minimal case gives the same tendencies of the interscale transport $tr^z_{ij}$ as observed in the reference case, including the inverse interscale transport of the Reynolds shear stress, despite the fact that the very-large-scale structures do not exist in this case. This suggests that the interscale transfers observed through the spanwise-Fourier-mode analysis may not represent the effect of the interaction between the near-wall and very-large-scale structures. This is further discussed in Sec.~\ref{sec:trz}.

\section{Discussions}

\subsection{On the relation between the spectral Reynolds-stress transport and \\the self-sustaining process of coherent structures}
\label{sec:ssp}

In Sec.~\ref{sec:lx}, we have shown that the streamwise length scales split mainly into two wavelength ranges: the larger $\lambda_x$ range $\lambda_x^+>400$ in which the energy is produced by $pr^x_{uu}$ and taken by the interscale energy transfer $tr^x_{uu}$, and the smaller $\lambda_x$ range $\lambda_x^+<400$ where the energy is supplied from larger $\lambda_x$ by $tr^x_{uu}$ and redistributed by the pressure-strain correlation $\pi^x_{uu}$. The boundary between these two $\lambda_x$ ranges has been found to be nearly constant throughout the channel and coincide with the streamwise-minimal domain length $L_x^+ \approx 400$. In the streamwise-minimal domain case, the streamwise wavelengths in the energy-producing (and -donating) $\lambda_x$ range are accounted for only by the streamwise-independent (i.e.~$k_x=0$) mode due to the limited degree of freedom, but otherwise the spectral energy transports observed in the reference case are basically retained, and thus the streamwise-minimal case reproduces the reference-case results fairly well. Given that both the inner and outer structures are retained in the streamwise-minimal case as shown in figures~\ref{fig:instAzy} and \ref{fig:ezuu}, these observations indicate that the energy production at very large streamwise wavelengths ($k_x\approx0$), the interscale energy transfer, and the energy redistribution at small wavelengths ($\lambda_x^+<400$) are the essential energy transport processes for both the inner and outer structures. 

In addition to the spectral energy transports described above, it should be further noted that the turbulent energy redistributed from $\ave{u^2}$ to $\ave{v^2}$ results in the production of the Reynolds shear stress $P_{-uv}=\ave{v^2} \mathrm{d}U/\mathrm{d}y$, which further leads to the reproduction of the streamwise turbulent energy $\ave{u^2}$ as $P_{uu}=-2\ave{uv} \mathrm{d}U/\mathrm{d}y$. Such a closed-loop transport of the Reynolds stresses consisting of production, interscale transfer, and inter-component redistribution resembles the scenario of the self-sustaining cycle of wall turbulence~\citep{hamilton_1995, waleffe_1997}: the significant $\ave{u^2}$ production on the nearly-$x$-independent modes is compatible well with the generation of the $x$-independent $u$-streaks; the interscale energy transfer by $tr^x_{uu}$, the streak instabilities; and the energy redistribution from $\ave{u^2}$ to $\ave{v^2}$ and $\ave{w^2}$ at relatively small $\lambda_x$, the generation of vortical structures from the streaks through their breakdown. While the self-sustaining process was originally proposed as the generating mechanism of the near-wall structures, recent studies~\citep[e.g.][]{hwang_2010,hwang_2011,rawat_2015,hwang_2016b,giovanetti_2017,cossu_2017} have provided overwhelming evidences indicating that the large-scale structures in the logarithmic and outer layers are also essentially maintained by similar self-sustaining cycle, rather than by the influence of the smaller-scale structures near the wall. Now we conjecture that the spectral energy transports observed in Sec.~\ref{sec:lx} are closely related with each subprocess of the self-sustaining cycle of the inner and outer structures, and further examine this conjecture in detail in the following.

An insight supporting the correspondence between the streak instabilities and the interscale energy transport $tr^x_{uu}$ is obtained by decomposing $tr^x_{uu}$. The interscale transport $tr^x_{ij}$ in the equation~(\ref{eq:exij}) is obtained as 
\begin{eqnarray}
tr^x_{ij}(y,k_x) = - \pd{Tr^x_{ij} (y,k_x)}{k_x},
\end{eqnarray}
where $Tr^x_{ij}$ represents the flux of the Reynolds stress in the streamwise wavenumber direction across $k_x$ from larger to smaller scale side, which is defined as 
\begin{eqnarray}
Tr^x_{ij} = -\ave{u^S_i u^S_k \pd{u^L_j}{x_k}} -\ave{u^S_j u^S_k \pd{u^L_i}{x_k}}
                 +\ave{u^L_i u^L_k \pd{u^S_j}{x_k}} +\ave{u^L_j u^L_k \pd{u^S_i}{x_k}}.
\end{eqnarray} 
Here, $u^L_i$ and $u^S_i$ are the large- and small-scale part of the fluctuating velocity $u_i$, respectively, obtained by applying low- and high-pass spatial filterings based on the streamwise Fourier mode with the cutoff wavenumber $k_x$ (see Appendix~\ref{app:a} for details). Then, the interscale flux of the streamwise turbulent energy $Tr^x_{uu}$ is expressed as
\begin{eqnarray}
Tr^x_{uu} &=& \underbrace{-2 \ave{u^S u^S \pd{u^L}{x}}}_{Tr^{x,1}_{uu}}  - \underbrace{2 \ave{u^S v^S \pd{u^L}{y}}}_{Tr^{x,2}_{uu}}  \underbrace{-2  \ave{u^S w^S \pd{u^L}{z}}}_{Tr^{x,3}_{uu}}  \nonumber \\
&& \hspace{2cm}  \underbrace{+2 \ave{u^L u^L \pd{u^S}{x}}}_{Tr^{x,4}_{uu}}  + \underbrace{2 \ave{u^L v^L \pd{u^S}{y}}}_{Tr^{x,5}_{uu}}  + \underbrace{2 \ave{u^L w^L \pd{u^S}{z}} }_{Tr^{x,6}_{uu}}, \label{eq:fluxuu}
\end{eqnarray}
and therefore the interscale transport $tr^x_{uu}$ is also decomposed as
\begin{eqnarray}
tr^x_{uu} = tr^{x,1}_{uu} + tr^{x,2}_{uu} + tr^{x,3}_{uu} + tr^{x,4}_{uu} + tr^{x,5}_{uu} + tr^{x,6}_{uu}, \label{eq:trxuu}
\end{eqnarray}
accordingly to $Tr^x_{uu}$. 

\begin{figure}
\begin{center}
	\includegraphics[width=1\hsize]{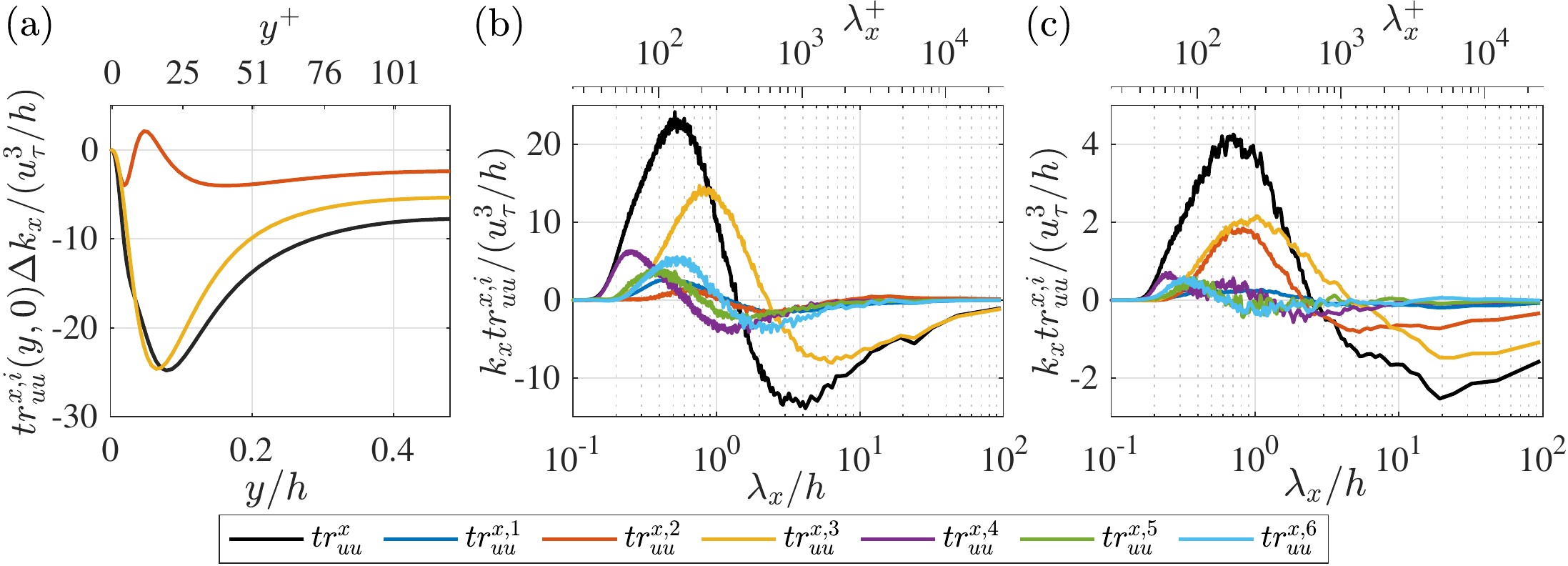}
	\caption{Comparison of various terms on the right-hand side of (\ref{eq:fluxuu}): (a) in the streamwise-minimal (A3) case, on the $x$-independent mode across the channel; (b, c) in the reference case, (b) at the near-wall location $y^+=16$ ($y/h=0.063$) and (c) at the channel centre $y/h=0.5$ across the whole $\lambda_x$ range investigated.}
	\label{fig:trxuui}
\end{center}
\end{figure}

Figure~\ref{fig:trxuui}(a) compares the contributions of two major terms on the right-hand side of (\ref{eq:trxuu}) to the energy transfer on the $x$-independent mode $tr^x_{uu}(y,0)$ in the streamwise-minimal case. Note that only $tr^{x,2}_{uu}(y,0)$ and $tr^{x,3}_{uu}(y,0)$ are presented here because the other terms are all zero, as explained in detail in Appendix~\ref{app:b}. As shown here, the energy is transferred from the $x$-independent mode mostly due to $tr^{x,3}_{uu}$, which represents the effect of the spanwise velocity gradient by the larger structure $\partial u^L/\partial z$, as one can see from (\ref{eq:fluxuu}). This corresponds to the spanwise variation of $u$ induced by the $x$-independent structures. In the reference case, on the other hand, the energy are taken from finite wavelengths unlike in the streamwise-minimal case, and therefore all terms in the right-hand side of (\ref{eq:trxuu}) can contribute to the interscale energy transfer from the large $\lambda_x$ range. The contributions of all terms in the reference case are presented for a near-wall location $y^+=16$ and the channel centre in figures~\ref{fig:trxuui}(b) and (c), respectively. As shown in figure~\ref{fig:trxuui}(b), the contribution by the third term $tr^{x,3}_{uu}$ is dominant at the near-wall location in the reference case at relatively large wavelengths ($\lambda_x^+>400$). In particular, the energy transfer form very large wavelengths $\lambda_x^+>1000$ is thoroughly due to $tr^{x,3}_{uu}$, which is consistent to the observations in the streamwise-minimal case with figure~\ref{fig:trxuui}(a). The other terms have roughly same magnitude and all become comparable to $tr^{x,3}_{uu}$ at relatively small wavelengths around $\lambda_x^+ \approx 100$. The energy transferred from large to middle $\lambda_x$ ranges by $tr^{x,3}_{uu}$ is further transferred towards smaller $\lambda_x$ range by the other terms, which may represent the turbulent energy cascade. The third term $tr^{x,3}_{uu}$ also plays a primary role in the interscale energy transport at the channel centre as shown in figure~\ref{fig:trxuui}(c). While the second term $tr^{x,2}_{uu}$ also shows a significant contribution, the contribution by $tr^{x,3}_{uu}$ is dominant at very larger wavelengths  $\lambda_x/h > 10$. Such dominant contributions by $tr^{x,3}_{uu}$ at large $\lambda_x$ are consistent with the observations by earlier studies that the main mechanism of the streak instabilities is inflectional instabilities due to the spanwise variation of $u$ induced by the streaks~\citep[e.g.][]{swearingen_1987,hamilton_1995,jimenez_1999}, and thus support our claim that $tr^x_{uu}$ in the relatively large $\lambda_x$ range mainly is closely related with the streak instabilities in the self-sustaining process both in the near-wall and in the central region of the channel.

\begin{figure}
\begin{center}
	\includegraphics[width=0.9\hsize]{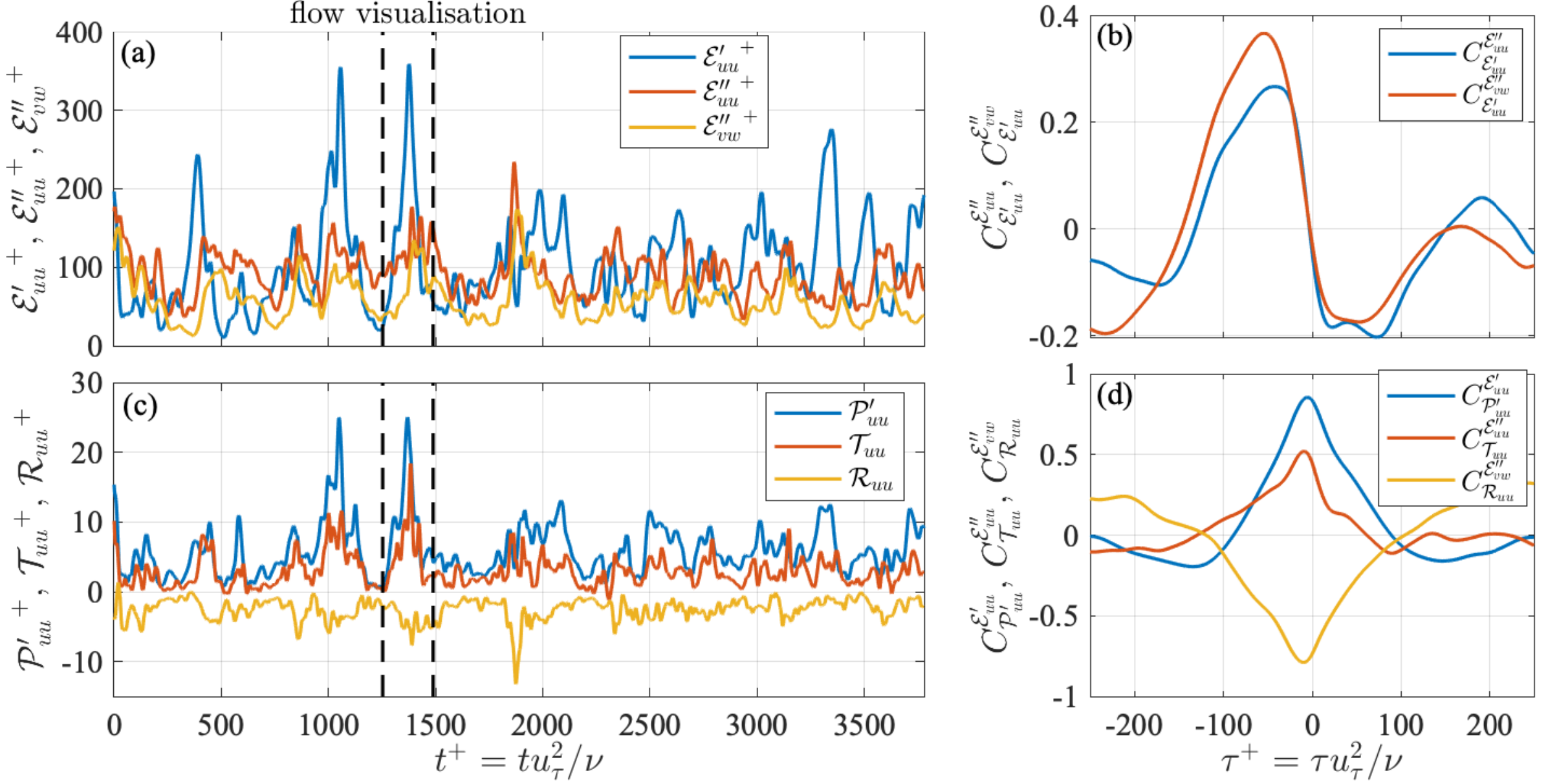}
	\caption{Time series and cross-correlation functions of integrated turbulent energies and Reynolds stress budgets obtained in Case~C: (a) time series of (blue) $\mathcal{E}^\prime_{uu}$, (red) $\mathcal{E}^{\prime \prime}_{uu}$, and (yellow)~$\mathcal{E}^{\prime \prime}_{vw}$ scaled by $u_\tau \nu$; (b) cross-correlation functions (blue) $C_{\mathcal{E}^\prime_{uu}}^{\mathcal{E}^{\prime \prime}_{uu}}$ and (red) $C_{\mathcal{E}^\prime_{uu}}^{\mathcal{E}^{\prime \prime}_{vw}}$; (c) time series of (blue) $\mathcal{P}^\prime_{uu}$, (red) $\mathcal{T}_{uu}$, and (yellow) $\mathcal{R}_{uu}$ scaled by $u_\tau^3$; (d) cross-correlation functions (blue) $C_{\mathcal{E}^\prime_{uu}}^{\mathcal{P}^\prime_{uu}}$, (red) $C_{\mathcal{E}^{\prime \prime}_{uu}}^{\mathcal{T}_{uu}}$, and (yellow) $C_{\mathcal{E}^{\prime \prime}_{vw}}^{\mathcal{R}_{uu}}$. The vertical black dashed lines in the panels~(a) and (c) indicate the time range of the flow visualisation presented in figure~\ref{fig:sspc}.}
	\label{fig:sspt}
\end{center}
\end{figure}

In order to more direclty examine the correspondence between the Reynolds stress budgets and the self-sustaining process, we investigate instantaneous flow fields obtained in Case~C, where both $L_x$ and $L_z$ are as small as their minimal sizes similarly to the flow configuration employed by \cite{hamilton_1995} to observe the self-sustaining cycle. The instantaneous velocity field is decomposed into the $x$-independent mode $u_i^\prime=\ave{u_i}_x$ and the $x$-dependent mode $u_i^{\prime \prime}=u_i-u_i^\prime$ (here $\ave{}_x$ represents spatial averaging in $x$-direction). Now we define the integrated turbulent energy on the $x$-independent mode as
\begin{eqnarray}
\mathcal{E}^\prime_{uu} = \int^{0.195h}_{0} \ave{{u^\prime}^2}_{xz} \mathrm{d}y,
\end{eqnarray}
and similarly the streamwise and lateral energies on the $x$-dependent modes as
\begin{eqnarray}
\mathcal{E}^{\prime \prime}_{uu} = \int^{0.195h}_{0} \ave{{u^{\prime \prime}}^2}_{xz} \mathrm{d}y, \quad
\mathcal{E}^{\prime \prime}_{vw} = \int^{0.195h}_{0} \left( \ave{{v^{\prime \prime}}^2}_{xz}+\ave{{w^{\prime \prime}}^2}_{xz}  \right) \mathrm{d}y,
\end{eqnarray}
and their time series are presented in figure~\ref{fig:sspt}(a). 
Here $\ave{}_{xz}$ indicates spatial averaging in $x$- and $z$-directions. The integration range $0 \leq y/h \leq 0.195$ ( $0 \leq y^+ \leq 40$ in the viscous unit) is chosen based on the $\ave{u^2}$ profile obtained in Case~C so that most of the turbulent energy related to the coherent structures near the bottom wall is included (see figure~\ref{fig:rs}). The time series of the integrated turbulent energies given in figure~\ref{fig:sspt}(a) present a typical periodic behaviour. It appears that they are well correlated with each other and the fluctuations of the $x$-dependent energies $\mathcal{E}^{\prime \prime}_{uu}$ and $\mathcal{E}^{\prime \prime}_{vw}$ follow that of the $x$-independent streamwise energy $\mathcal{E}^{\prime}_{uu}$ with a slight delay. The time delays are  quantified in figure~\ref{fig:sspt}(b), where the cross-correlation functions between $\mathcal{E}^{\prime}_{uu}$ and $\mathcal{E}^{\prime \prime}_{uu}$ and between $\mathcal{E}^{\prime}_{uu}$ and $\mathcal{E}^{\prime \prime}_{vw}$ defined as
\begin{eqnarray}
C_{\mathcal{E}^{\prime}_{uu}}^{\mathcal{E}^{\prime \prime}_{uu}} (\tau) = \frac{\ave{\mathcal{E}^{\prime}_{uu}(t) \mathcal{E}^{\prime \prime}_{uu} (t + \tau)}}{\sqrt{\ave{{\mathcal{E}^{\prime}_{uu}}^2}} \sqrt{\ave{{\mathcal{E}^{\prime \prime}_{uu}}^2}}}, \quad
C_{\mathcal{E}^{\prime}_{uu}}^{\mathcal{E}^{\prime \prime}_{vw}} (\tau) = \frac{\ave{\mathcal{E}^{\prime}_{uu}(t) \mathcal{E}^{\prime \prime}_{vw} (t + \tau)}}{\sqrt{\ave{{\mathcal{E}^{\prime}_{uu}}^2}} \sqrt{\ave{{\mathcal{E}^{\prime \prime}_{vw}}^2}}},\label{eq:crscor}
\end{eqnarray} 
are presented. As can be seen here, the correlation peaks of both cross-correlation functions are located at negative $\tau$ around $\tau^+\approx-50$, and $C_{\mathcal{E}^{\prime}_{uu}}^{\mathcal{E}^{\prime \prime}_{vw}}$ gives larger delay than $C_{\mathcal{E}^{\prime}_{uu}}^{\mathcal{E}^{\prime \prime}_{uu}}$. This indicates that the fluctuation of the $x$-independent streamwise turbulent energy $\mathcal{E}^\prime_{uu}$ is followed by that of $x$-dependent energy $\mathcal{E}^{\prime \prime}_{uu}$, and they are further followed by the lateral turbulent energy $\mathcal{E}^{\prime \prime}_{vw}$. Such an observed sequence of $\mathcal{E}^\prime_{uu}$, $\mathcal{E}^{\prime \prime}_{uu}$, and $\mathcal{E}^{\prime \prime}_{vw}$ indicates that they mainly represent the energies of $x$-independent streaks, wavy streaks and vortical structures of the self-sustaining process, respectively.

Similarly, the integrated turbulent energy production on the $x$-independent mode $\mathcal{P}^\prime_{uu}$, interscale energy flux from the $x$-independent towards -dependent modes $\mathcal{T}_{uu}$, and pressure-strain energy redistribution $\mathcal{R}_{uu}$ are defined as
\begin{subeqnarray}
\mathcal{P}^\prime_{uu} &=& \int^{0.195h}_0 \ave{\left. \widetilde{pr^x_{uu}} \right | _{k_x=0} \Delta k_x}_{xz} \mathrm{d}y = \int^{0.195h}_0 - 2 \ave{u^\prime v^\prime}_{xz} \frac{\mathrm{d}U}{\mathrm{d}y} \mathrm{d}y, \\
\mathcal{T}_{uu} &=& \int^{0.195h}_0 \ave{\left. \widetilde{Tr^{x,3}_{uu}} \right|_{k_x=0}}_{xz} \mathrm{d}y = \int^{0.195h}_0  -2 \ave{u^{\prime \prime} w^{\prime \prime} \pd{u^\prime}{z}}_{xz} \mathrm{d}y,  \\
\mathcal{R}_{uu} &=& \int^{0.195h}_0 \ave{\widetilde{\Pi_{uu}}}_{xz} (y) \mathrm{d}y =\int^{0.195h}_0  2 \ave{p^{\prime \prime} \pd{u^{\prime \prime}}{x}}_{xz} \mathrm{d}y,  
\end{subeqnarray}
where the tilde~$\widetilde{\:}$ indicates the instantaneous values of the spectral budget term that are not averaged in $x$- or $z$-direction or in time. The times series of these integrated budget terms are presented in figure~\ref{fig:sspt}(c). These time series also indicate certain periodic behaviours and are significantly correlated with each other, similarly to those of the integrated turbulent energies in the panel~(a). Figure~\ref{fig:sspt}(d) presents cross-correlation functions $C_{\mathcal{P}^{\prime}_{uu}}^{\mathcal{E}^{\prime}_{uu}}$, $C_{\mathcal{T}_{uu}}^{\mathcal{E}^{\prime \prime}_{uu}}$, and $C_{\mathcal{R}_{uu}}^{\mathcal{E}^{\prime \prime}_{vw}}$ defined in the same manner as in equations~(\ref{eq:crscor}). As shown here, all three cross-correlation functions present the maximum magnitude of correlation located at negative $\tau$ on the order of $\tau^+ \approx 10$. This indicates that $\mathcal{E}^\prime_{uu}$, $\mathcal{E}^{\prime \prime}_{uu}$, and $\mathcal{E}^{\prime \prime}_{vw}$ response immediately to increase/decrease of the energy production $\mathcal{P}^\prime_{uu}$, the interscale energy transfer $\mathcal{T}_{uu}$, and the pressure-strain energy redistribution $\mathcal{R}_{uu}$, respectively. In particular, the peak magnitudes of these cross-correlation functions are all significantly high (more than 0.5), which also indicates the close causal relationship between these turbulent energies and the Reynolds stress budget terms.

\begin{figure}
\begin{center}
	\includegraphics[width=0.9\hsize]{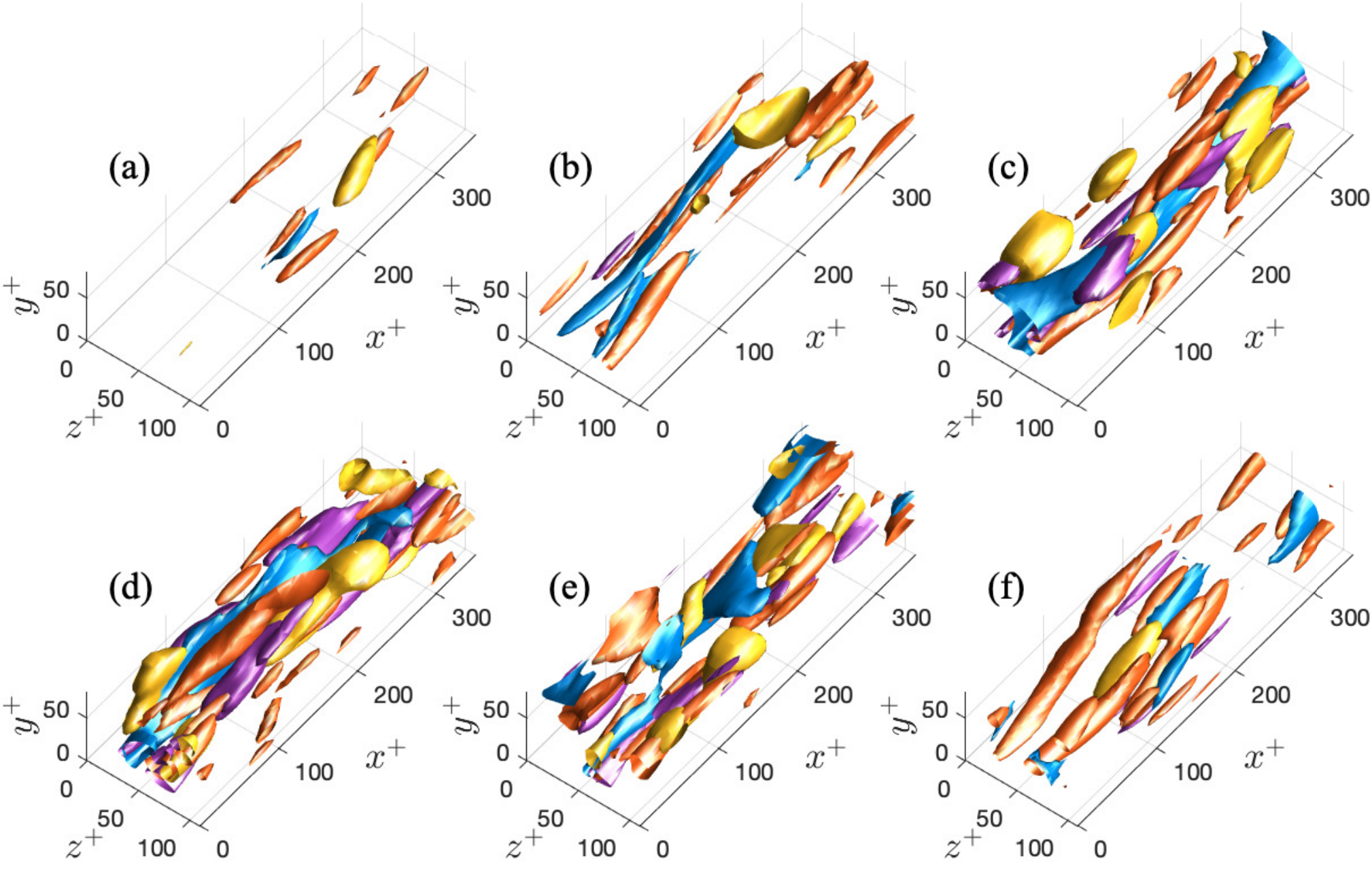}
	\caption{Visualisations of instantaneous flow field in the near region of the bottom wall ($0 \leq y^+ \leq 60$) observed in Case~C at (a)~$t^+=1253$, (b)~$t^+=1293$, (c)~$t^+=1370$, (d)~$t^+=1386$, (e)~$t^+=1400$, and (f)~$t^+=1489$. Full range of the computational domain is shown in $x$- and $z$-directions. The 3D surfaces of different colours in each panel represent iso-surfaces of (blue)~$u^+=-5$, (red)~$Q^+=-0.04$, (yellow)~$\tilde{\Pi}_{uu}^+=-0.75$, and (purple)~$\left. \widetilde{Tr_{uu}^{x,3}}^+ \right|_{k_x=0} =1.45$. The flow field shown is low-pass filtered at the cutoff wavenumber $k_{x,c}h/2\pi=1.88$ ($\lambda_{x,c}^+=118$) for clear visualisation of coherent structures. }
	\label{fig:sspc}
\end{center}
\end{figure}

Figure~\ref{fig:sspc} presents flow visualisations of instantaneous flow field at six different instances during the period indicated by black dashed lines in figures~\ref{fig:sspt}(a) and (c). In each panel, the near region of the bottom wall, $0 \leq y^+ \leq 60$, is visualised and the colour surfaces represent iso-surfaces of (blue) $u^+=-5$, (red) $Q^+=-0.04$, (yellow) $\widetilde{\Pi}_{uu}^+=-0.75$, and (purple)~$\left. \widetilde{Tr_{uu}^{x,3}}^+ \right|_{k_x=0}=1.45$, where $Q$ is the second invariant of the velocity-gradient tensor. The instantaneous flow fields shown here are low-passed filtered with a cutoff wavelength $k_{x,c}h/2\pi=1.88$ ($\lambda_{x,c}^+=118$) for clear visualisations of coherent structures, but most of the turbulent energies (more than 90\% of $\ave{u^2}$ and 68\% of $\ave{v^2}+\ave{w^2}$) are still retained in the flow field visualised here. In panels~(a)-(c) (from $t^+=1293$ to 1370), an evolution of a low-speed streak is observed and streamwise-elongated vortices are also found aligned with the streak. Particularly, in the panel~(c) the streak has grown enough to penetrate the computational domain in $x$-direction. This instance ($t^+=1370$) actually corresponds to the peak of the $x$-independent streamwise turbulent energy $\mathcal{E}^\prime_{uu}$ in figure~\ref{fig:sspt}(a), and the $x$-dependent streamwise turbulent energy $\mathcal{E}^{\prime \prime}_{uu}$ is also increasing at this instance. Therefore the streak visualised in figure~\ref{fig:sspc}(c) is somewhat wavy. It can also be observed that the regions of significant interscale energy transfer from $x$-independent to -dependent modes (purple) and inter-component energy transfer by pressure-strain correlation (yellow) start to appear around the streak. The panel~(d) (at $t^+=1386$) corresponds to the peak of the interscale energy transfer $\mathcal{T}_{uu}$ in figure~\ref{fig:sspt}(c) and at this instance the $x$-dependent energy $\mathcal{E}^{\prime \prime}_{uu}$ in figure~\ref{fig:sspt}(a) is also saturated. Hence, the low-speed streak visualised in figure\ref{fig:sspc}(d) is strongly wavy and surrounded by the regions of significant interscale energy transfer. It is also observed that the pressure-strain energy redistribution is significant and streamwise-elongated vortices visualised by $Q$ are evolving around the wavy streak. In the panel~(e) (at $t^+=1400$), the streak is shown to have already broken down. However, the regions of significant pressure-strain energy redistribution are still found and vortical structures also remain. As time further proceeds such vortical structures further evolve and remain after the low-speed streaks become hardly visible as shown in the panel~(f), which eventually give rise to next streaks.

\begin{figure}
\begin{center}
	\includegraphics[width=0.9\hsize]{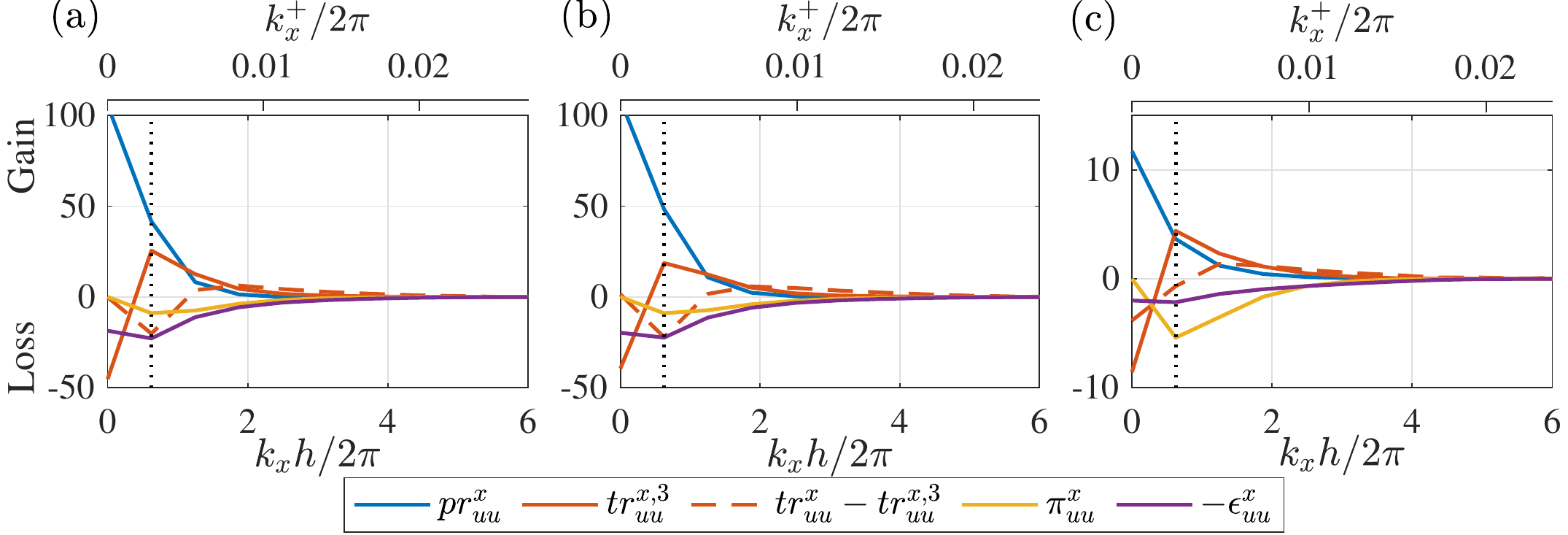}
	\caption{Spectral budget of streamwise turbulent energy transport: (a) at $y^+=16$ in Case~C; (b) at $y^+ = 16$ in Case~A3; at channel centre in Case~A3. In each panel different lines represent (blue, solid) production $pr^x_{uu}$, (red, solid) interscale transport $tr^{x,3}_{uu}$, (red, dashed) contribution from other terms of the interscale transport than $tr^{x,3}_{uu}$, (yellow) pressure-strain energy redistribution $\pi^x_{uu}$, and (purple) viscous dissipation $-\epsilon^x_{uu}$. The values are scaled by $u_\tau^3$. The vertical black dotted lines indicate $k_xh/2\pi=1/1.6$, which corresponds to the streamwise domain size in Cases~A3 and C.}
	\label{fig:ca3}
\end{center}
\end{figure}

The above observations in Case~C of the growth/breakdown of a streak and vortices in relation to the turbulent energy budget terms agree fairly well with the scenario of the self-sustaining cycle, and thus support our conjecture about their close relations. 
The detailed balance between the spectral budget terms in Case~C is presented in figure~\ref{fig:ca3}(a) for a near-wall location $y^+ = 16$. As shown here the peak of the energy production is located at $k_x=0$ accounting for nearly 70\% of the overall energy production. The interscale energy transport $tr^{x,3}_{uu}$ is shown to dominate the energy transport at small wavenumbers near $k_x=0$, transferring nearly half the energy produced at $k_x=0$ mainly to the smallest non-zero wavenumber $k_x =2\pi/1.6h$, where the most significant energy redistribution by $\pi^x_{uu}(y,k_x)$ is found. The viscous dissipation $-\varepsilon^x_{uu}$ is additionally presented together, shown to dissipate the energy broadly throughout the investigated wavenumber range. As the flow fields visualised in figure~\ref{fig:sspc} account for most of $k_x$ at which $pr^x_{uu}$, $tr^x_{uu}$, and $\pi^x_{uu}$ are significant in their profiles given in figure~\ref{fig:ca3}(a), it is reasonably interpreted that the spectral budget balance presented in figure~\ref{fig:ca3}(a) mainly represents the energy transport related to the dynamics of the streaks and vortices observed in figure~\ref{fig:sspc}, which resembles the self-sustaining cycle.

The spectral energy budgets obtained in the streamwise-minimal (A3) case also give similar tendencies as observed in the Case~C. Figures~\ref{fig:ca3}(b) and (c) present the profiles of the spectral budget terms at a near-wall location $y^+=16$ and at the channel centre in the streamwise-minimal case, respectively. Both budget balances indicate the maximum energy production at $k=0$, the energy transfer from $k_x=0$ to $k_x=2\pi/1.6h$ mostly by $tr^{x,3}_{uu}$, and the most significant energy redistribution by $\pi^x_{uu}$ at $k_x=2\pi/1.6h$, similarly to the budget balance in figure~\ref{fig:ca3}(a). From this observation one can infer that these spectral budget terms in the near-wall and central regions of the channel in the streamwise-minimal case represent the energy transports related to the self-sustaining cycle of the near-wall and very-large-scale structures, respectively.

Such a connection between the self-sustaining cycle and the Reynolds stress transports observed through the spectral analysis was already pointed out by \cite{cho_2018}. Their discussion was based on the spanwise spectra of the turbulent energy production and the energy redistribution by the pressure-strain correlation. They showed that the peak location in the distributions of these spectra follows the scaling $\lambda_z \sim 5 y$, which is consistent with the attached eddy hypothesis, and thereby conjectured that the energy production and the pressure-strain correlation respectively correspond to the streak generation and the regeneration of the streamwise vortices in the self-sustaining process of each wall-attached eddy. In the present study, we have investigated the spectral transport of the Reynolds stress based on the streamwise-Fourier-mode analysis with reduced-size computational domains, and have deduced a similar interpretation of the major terms of the spectral transport equation of the turbulent energy. 

\subsection{Inverse interscale energy transport in the regeneration of the streamwise vortices\\ in the self-sustaining process} 

\begin{figure}
\begin{center}
	\includegraphics[width=0.85\hsize]{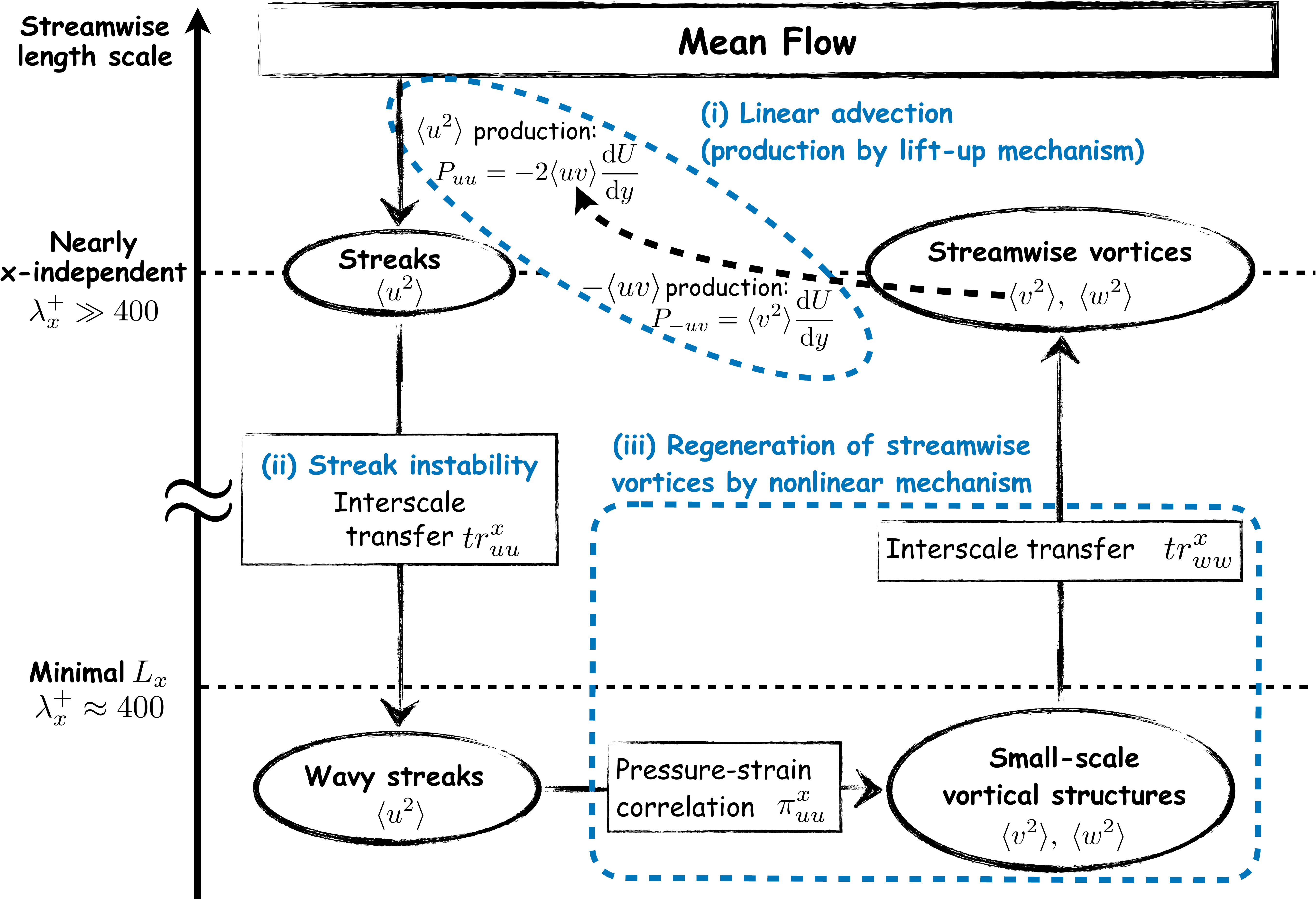}
	\caption{Schematics of the closed-loop spectral transport of the Reynolds stresses with streamwise length scales. The subprocesses of self-sustaining cycle that likely correspond are also shown together. The solid black arrows represent the flow of energy transfer, and the dashed solid lines connecting $\ave{v^2}$ and $\ave{uv}$ in $P_{uu}$ means that $\ave{v^2}$ triggers $P_{uu}$ through $P_{-uv}$. }
	\label{fig:ssp}
\end{center}
\end{figure}

The closed-loop transport of the Reynolds stresses is schematically summarised in figure~\ref{fig:ssp} with their streamwise length scales, and the subprocesses of the self-sustaining cycle that likely correspond are also included together. From figure~\ref{fig:ssp} one can find an interesting point regarding the reproduction of the streamwise turbulent energy~$\ave{u^2}$. As explained in the previous section, the turbulent energy production $pr^x_{uu}=2 E^x_{-uv} \mathrm{d}U/\mathrm{d}y$ results from the wall-normal turbulent energy $E^x_{vv}$ through the production of the Reynolds shear stress $pr^x_{-uv}=E^x_{vv} \mathrm{d}U/\mathrm{d}y$. What should be noted here is that both productions $pr^x_{-uv}$ and $pr^x_{uu}$ are linear processes that do not involve any scale interactions, which means that there has to exist a certain amount of the wall-normal turbulent energy $E^x_{vv}$ at large $\lambda_x$ in order that the production of $\ave{u^2}$ takes place. However, the energy source for $\ave{v^2}$ is the energy redistribution from $\ave{u^2}$, which, as observed in Sec.~\ref{sec:lx}, mainly occurs at relatively small $\lambda_x$ ($\lambda_x^+<400$) and is quite small (in the streamwise-minimal case, zero) in the energy-producing $\lambda_x$ range as $\partial u/\partial x$ is small at such large $\lambda_x$. Of course there is no direct energy production from the mean flow for $\ave{v^2}$, unlike for $\ave{u^2}$ at large scales. Hence, any typical energy supply to $\ave{v^2}$ is not found at large scales despite the fact that $\ave{v^2}$ at large scales is indispensable for the turbulent energy production. This indicates a certain energy transfer to $\ave{v^2}$ at large scales from smaller scales where the energy redistribution from $\ave{u^2}$ to $\ave{v^2}$ and $\ave{w^2}$ is significant.

\begin{figure}
\begin{center}
	\includegraphics[width=1\hsize]{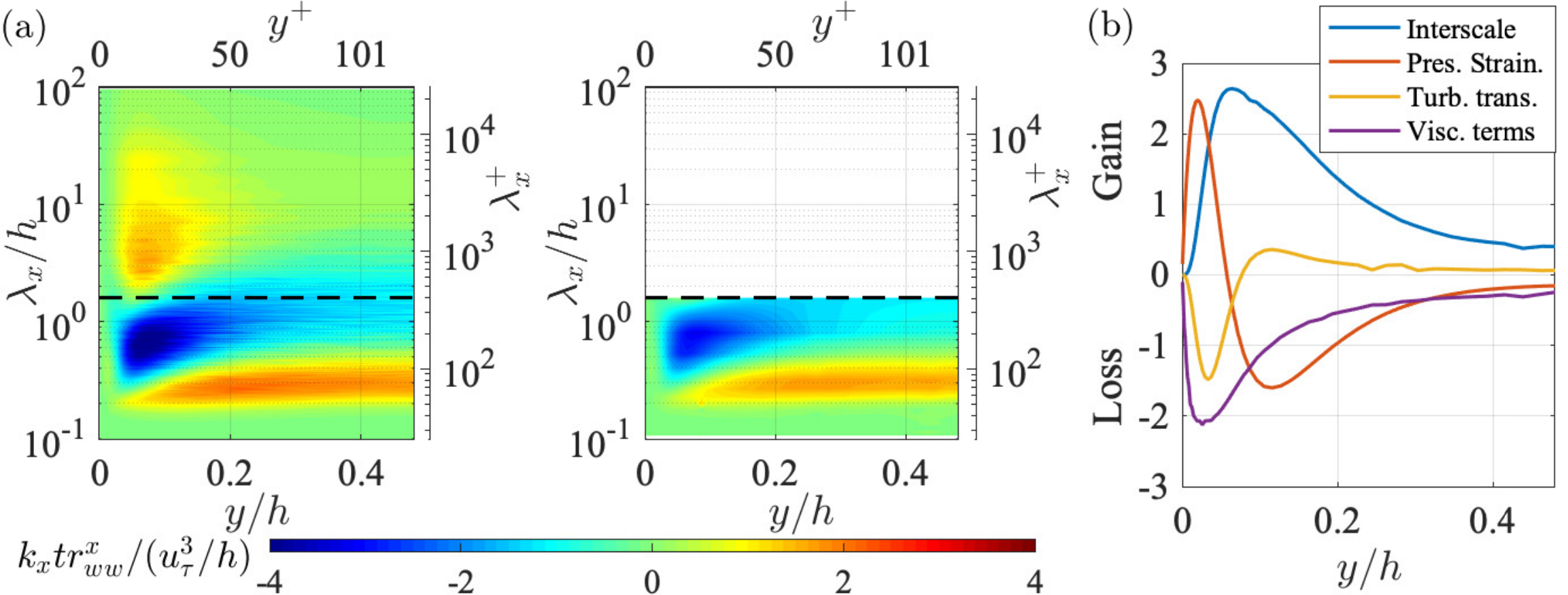}
	\caption{(a) Space-wavelength ($y$-$\lambda_x$) diagrams of the premultiplied interscale transport of the spanwise turbulent energy in the streamwise wavenumber direction $k_x tr^x_{ww}$ obtained in the reference and streamwise-minimal (A3) cases. The horizontal black dashed lines represent the streamwise domain size in the streamwise-minimal case $\lambda_x/h=1.6$. (b)~The spectral budget balance of $\ave{w^2}$ transport at $k_x=0$ in the streamwise-minimal case: (blue)~the interscale transport $tr^x_{ww}(y,0)$, (red)~the pressure-strain cospectrum $\pi^x_{ww}(y,0)$, (yellow)~the turbulent spatial transport $d^{t,x}_{ww}(y,0)$, (purple)~the viscous terms $d^{\nu,x}_{ww}(y,0)+\epsilon^x_{ww}(y,0)$. The values are scaled by $u_\tau^3/h$.}
	\label{fig:trxww}
\end{center}
\end{figure}

It is, in fact, the inverse interscale energy transfer of $\ave{w^2}$ and the energy redistribution from $\ave{w^2}$ to $\ave{v^2}$ that provides energy to the $\ave{v^2}$ component at large $\lambda_x$. As shown in figure~\ref{fig:trxww}(a), the interscale transport $tr^x_{ww}$ in the reference case clearly indicates the energy transfer from smaller to larger $\lambda_x$ particularly in the near-wall region, as well as the forward (from larger to smaller $\lambda_x$) transfer, and this tendency is also reproduced in the streamwise-minimal case as the transfer from the $\lambda_x^+<400$ range to the $x$-independent mode. The interscale transfer of the wall-normal turbulent energy $tr^x_{vv}$ basically indicates forward interscale transfers, but is one-order smaller in magnitude than the reversed energy transfer by $tr^x_{ww}$ (not shown). Such reversed energy cascade in the near-wall region has been repeatedly reported in earlier studies of turbulent channel flow~\cite[e.g.][]{saikrishnan_2012,cimarelli_2013,cimarelli_2016,cho_2018,hamba_2019,lee_2019}, while their roles in maintaining flow structures are still unclear. Figure~\ref{fig:trxww}(b) presents the $\ave{w^2}$ budget of the $x$-independent mode in the streamwise-minimal case. As shown here, the main energy source for $\ave{w^2}$ on this mode is the inverse energy transfer from $k_x>0$ throughout the channel and the turbulent spatial transport $d^t_{ww}$ carries the energy from the near-wall to the channel central region. The pressure-strain correlation $\pi^x_{ww}(y,0)$ represents the energy exchange between $\ave{v^2}$ and $\ave{w^2}$ as $\pi^x_{uu}=0$ and $\pi^x_{vv}=-\pi^x_{ww}$ at $k_x=0$, and the profile of $\pi^x_{ww}(y,0)$ indicates that the energy is transferred from $\ave{w^2}$ to $\ave{v^2}$ in the central region of the channel while it is transferred in the opposite direction in the near wall region. Integrating $tr^x_{ww}(y,0)$ and $\pi^x_{ww}(y,0)$ gives $\int^{0.5h}_{0} \pi^x_{ww}(y,0) \mathrm{d}y \approx -0.35 \int^{0.5h}_{0} tr^x_{ww}(y,0) \mathrm{d}y$, indicating that the net energy transfer from $\ave{w^2}$ to $\ave{v^2}$ by $\pi^x_{ww}$ at $k_x=0$ is about 35\% of the total energy gain from smaller scales by $tr^x_{ww}$. 

\begin{figure}
\begin{center}
	\includegraphics[width=0.9\hsize]{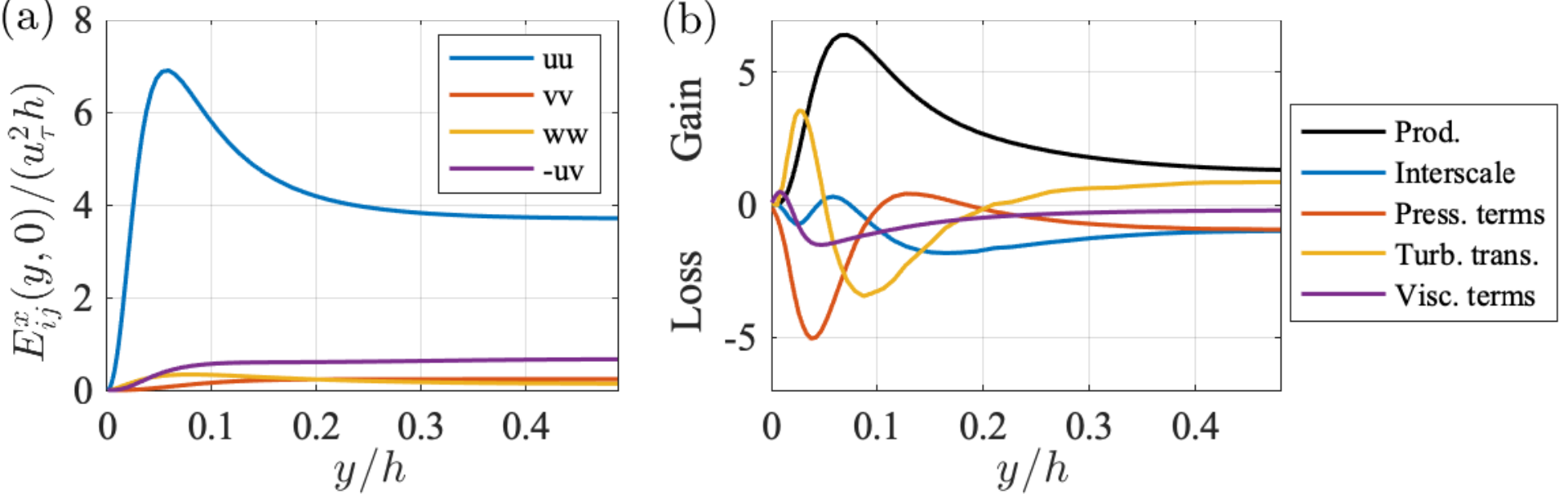}
	\caption{(a) Profiles of the Reynolds stress spectra $E^x_{ij}$ and (b)~the spectral budget balance of $-\ave{uv}$ transport at $k_x=0$ in the streamwise-minimal (A3) case. In the panel (b), the lines with different colours represent (black)~the production $pr^x_{-uv}(y,0)$, (blue)~the interscale transport $tr^x_{-uv}(y,0)$, (red)~the pressure-related terms $\pi^x_{-uv}(y,0)+d^{p,x}_{-uv}(y,0)$, (yellow)~the turbulent spatial transport $d^{t,x}_{-uv}(y,0)$, and (purple)~the viscous terms $d^{\nu,x}_{-uv}(y,0)+\epsilon^x_{-uv}(y,0)$, and the values are scaled by $u_\tau^3/h$.}
	\label{fig:vwuv}
\end{center}
\end{figure}

Due to such an energy supply to the wall-normal and spanwise turbulent energies at $k_x=0$, the energy spectra $E^x_{vv}(y,0)$ and $E^x_{ww}(y,0)$ show non-zero energy distributions as presented in figure~\ref{fig:vwuv}(a). Their magnitudes are small as compared to the streamwise component $E^x_{uu}(y,0)$ as shown here, since the energy amount supplied to $E^x_{vv}(y,0)$ and $E^x_{ww}(y,0)$ from smaller scales via $tr^x_{ww}$ is only about 5\% of the total energy amount produced from the mean flow at $k_x=0$ by $pr^x_{uu}$. The Reynolds shear stress cospectrum $E^x_{-uv}(y,0)$ is also presented in the figure, shown to be on the same magnitude as $E^x_{vv}(y,0)$ and $E^x_{ww}(y,0)$. The budget balance of $-\ave{uv}$ transport at $k_x=0$ presented in figure~\ref{fig:vwuv}(b) shows that the production term $pr^x_{-uv}(y,0)=E^x_{vv}(y,0) \mathrm{d}U/\mathrm{d}y$ is the main energy source throughout the channel. This indicates that the small amount of the wall-normal energy $E^x_{vv}(y,0)$ indeed maintains the Reynolds shear stress at $k_x=0$, which subsequently results in the reproduction of $\ave{u^2}$ on the $x$-independent mode via the turbulent energy production $pr^x_{uu}(y,0)=2 E^x_{-uv} (y,0) \mathrm{d}U/\mathrm{d}y$. In the reference case the turbulent energy production at large scales is also maintained by similar energy transports including the inverse interscale energy transfer, as described above (not shown). 

Thus, it has been shown that the turbulent energy production at large streamwise length scales is maintained by the inverse interscale transfer of spanwise turbulent energy from smaller scales. 
This also means that, supposing that these Reynolds-stress budget terms are closely related to the self-sustaining process as we describe in figure~\ref{fig:ssp}, the regeneration of the streamwise vortices through breakdowns of wavy streaks should include an inverse interscale energy transfer as well as the inter-component energy transfer by the pressure-strain correlation. In figure~\ref{fig:ssp}, we highlight this point in ``(iii)~Regeneration of streamwise vortices by nonlinear mechanism'' (see in the blue dashed square box) by denoting the structure generated from `Wavy streaks' through their breakdowns as `Small-scale vortical structures', distinguishing them from `Streamwise vortices', which directly lead to generation of `Streaks'. `Small-scale vortical structures' are generated at small scales ($\lambda_x^+<400$) through streak breakdowns by the pressure-strain energy redistribution, and `Streamwise vortices' is regenerated at large scales by the inverse interscale energy transfer, which eventually results in regeneration of `Streaks' by triggering turbulent energy production. 

\cite{hamba_2019} focused on the inverse interscale transport of $\ave{w^2}$ similarly to the present study and attempted to extract the related vortical structure by means of conditional averaging. He found a longitudinal streamwise vortex accompanied with a shorter vortex located upstream, which indicates that the interactions between these vortices are responsible to the inverse energy cascade. Although the connection between such structures found in his work and the regeneration process of the streamwise vortices in the self-sustaining cycle is still unclear, the inverse energy transfer of $\ave{w^2}$ towards large $\lambda_x$ observed in the present study may also be caused by similar interactions between the long (at large $\lambda_x$) and short ($\lambda_x^+<400$) vortices.   

\subsection{On the DNS with streamwise-minimal domain at higher Reynolds numbers}

As shown in Sec.~\ref{sec:lx}, the reason underlying the good agreement between the reference and streamwise-minimal cases is that the upper boundary of the $\lambda_x$ range where the turbulent energy is supplied from larger $\lambda_x$ via the interscale transport $tr^x_{uu}$ and redistributed by the pressure-strain correlation $\pi^x_{uu}$ is rather independent of $y$ and smaller than the streamwise minimal length $L_x=1.6h$ throughout the channel. As our investigations are limited in a low Reynolds number range $Re_\tau \approx 126$, it should be further examined whether or not this is still the case even at higher Reynolds numbers. Regarding this point one may refer to the works by \cite{abe_2018} and \cite{lee_2019}; \cite{abe_2018} performed DNSs of turbulent channel flows at $Re_\tau=1020$ comparing the results obtained with very-large and streamwise-minimal domains, and \cite{lee_2019} also carried out the DNS of a turbulent channel flow at $Re_\tau = 1000$ providing detailed data of the spectral transport of the Reynolds stresses. 

\begin{figure}
\begin{center}
	\includegraphics[width=1\hsize]{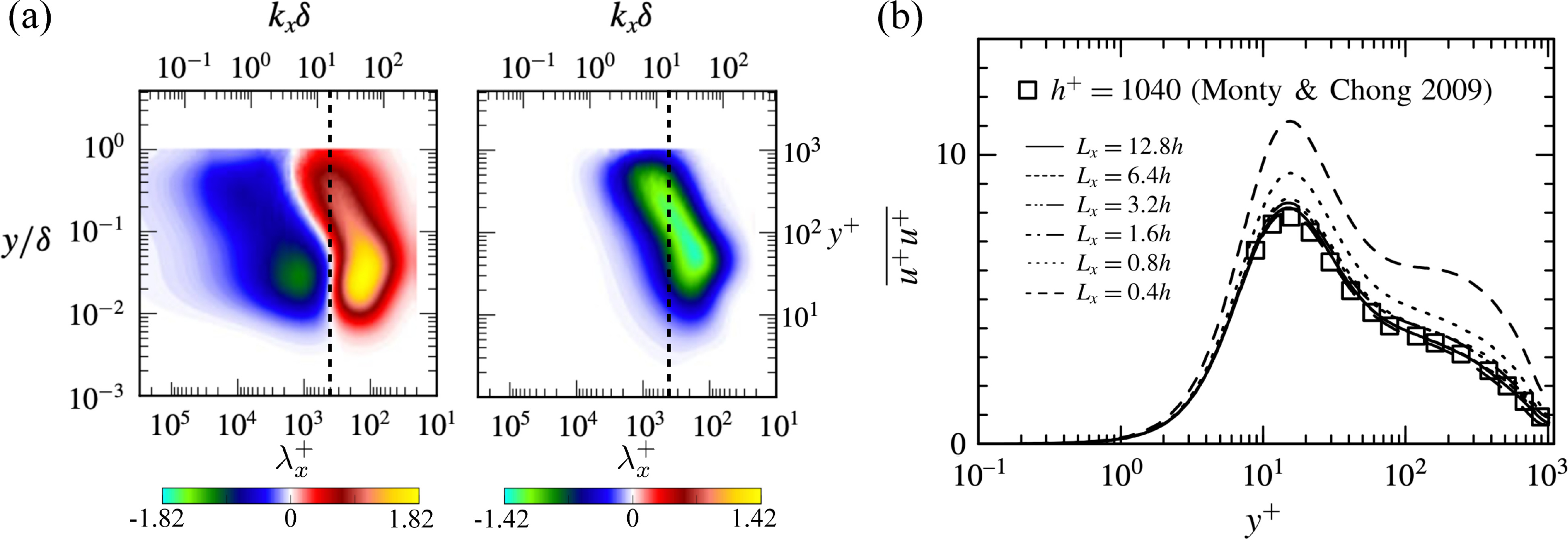}
	\caption{(a) $\lambda_x$-$y$ diagrams of (left) the interscale energy transport and (right) the pressure-strain cospectrum for $\ave{u^2}$, scaled by $u_\tau^4/\nu$, obtained in a turbulent channel flow at $Re_\tau=1000$, reproduced with permission from figures~26(a) and 13(b) in \cite{lee_2019} (the vertical dashed lines at $\lambda_x^+=400$ are added in the present study): (b) Profiles of streamwise turbulent intensity $\ave{u^2}/u_\tau^2$ obtained in a turbulent channel flow at $Re_\tau=1020$ with various streamwise domain length $L_x$, reproduced with permission from figure~7(a) in \cite{abe_2018}.}
	\label{fig:la}
\end{center}
\end{figure}

In figure~\ref{fig:la}(a), we reproduce from \cite{lee_2019} the $\lambda_x$-$y$ distributions of the premultiplied interscale transport of the streamwise turbulent energy in the streamwise wavenumber direction, denoted as ${E_{uu,x}^{T^{||}}}$ by their notations, and the pressure-strain cospectrum $\pi^x_{uu}$ in a turbulent channel flow at $Re_\tau=1000$. 
In these distributions, blue and red colours represent the energy loss and gain, respectively. In the ${E_{uu,x}^{T^{||}}}$ distribution (on the left of the panel~(a)), one can see that there exists a certain $y$ region up to about $y^+=100$ where the boundary between the energy-donating (blue) and -receiving (red) $\lambda_x$ ranges is hardly dependent on $y$ and coincides with $\lambda_x^+ \approx 400$ (see the vertical black dashed line), similarly to our observation in the turbulent plane Couette flow. Further away from the wall, however, the peak locations of the ${E_{uu,x}^{T^{||}}}$ distribution clearly vary depending on the distance from the wall, which appears to be proportional to $y$.  Similarly to this, it is also shown in the the right of the panel~(a) that the $\pi^x_{uu}$ distribution also follows the scaling with $y$ away from the wall about $y^+>100$. This means that the aforementioned observation in the present study that the distributions of $tr^x_{uu}$ and $\pi^x_{uu}$ are rather independent of $y$ is merely a low-Reynolds-number effect. 

As shown in the $\pi^x_{uu}$ distribution in figure~\ref{fig:la}(a), significant amount of the pressure-strain cospectrum is distributed in the range $\lambda_x^+>400$ for $y^+>100$ at the Reynolds number $Re_\tau=1000$, which cannot be taken into account in the simulation with the streamwise-minimal domain. This would result in a significant underestimation of the energy redistribution from $\ave{u^2}$ to the other components, and consequently $\ave{u^2}$ would be remarkably overestimated. This is indeed observed in figure~\ref{fig:la}(b), which presents the profiles of $\ave{u^2}$ in turbulent channel flows at a similar Reynolds number $Re_\tau=1020$ computed with different streamwise domain length $L_x$, reproduced from \cite{abe_2018}. As shown here, as $L_x$ decreases the peak magnitude of $\ave{u^2}^+$ increases, consistently to the observation in the present study, and in the case of $L_x=0.4h$, i.e.~$L_x^+\approx 400$, the overestimation of $\ave{u^2}$ magnitude is significant throughout the channel. This tendency is particularly remarkable in the logarithmic region around $y^+ \approx 100$; a plateau of $\ave{u^2}$ in this $y^+$ range, which would correspond to the outer peak of $\ave{u^2}$ at higher Reynolds numbers, is already observed, whereas in larger $L_x$ cases it does not emerge. This is likely because in the streamwise-minimal ($L_x=0.4h$) case the significant contribution by $\pi^x_{uu}$ in $L_x^+>400$ is excluded. Hence, it is expected that $\ave{u^2}$ would be similarly overestimated (and correspondingly $\ave{v^2}$ and $\ave{w^2}$ underestimated) particularly in relatively far-wall region $y^+>100$ in the DNS of turbulent plane Couette flow at higher Reynolds numbers with the streamwise-minimal domain.

\subsection{On the inverse interscale transfer of the Reynolds stresses observed through one-dimensional spanwise-Fourier-mode analysis}
\label{sec:trz}

As observed in Sec.~\ref{sec:lz}, the distribution of the interscale transport of the Reynolds shear stress $tr^z_{-uv}$ obtained in the spanwise-minimal (B2) case indicates the transfer from smaller to larger $\lambda_z$ throughout the channel similarly to the reference-case result, despite the fact that the very-large-scale structure in the channel core region does not exist in Case~B2 due to the small $L_z$. This suggests that the interscale Reynolds-stress transfers observed through one-dimensional spanwise-Fourier-mode analysis may not represent the interaction between the inner- and outer-layer structures, unlike as suggested by \cite{kawata_2018}. Furthermore, even in Case~C (both $L_x$ and $L_z$ as small as minimal), the same tendencies of $tr^z_{ij}$ are still retained (not shown). This indicates that the spanwise interscale transport $tr^z_{ij}$ may rather be related to the individual dynamics of each near-wall and very-large-scale structure, such as their self-sustaining process. 

A Fourier mode interaction that might possibly be related with the inverse interscale transfer of the Reynolds shear stress can be found in the work by \cite{hamilton_1995}. In their investigation on the temporal variations of the energies of different Fourier modes during the self-sustaining cycle (in their figure~3), it is shown that when the energy of the $x$-independent streak, $M(0,\beta)$ by their notation ($\beta$ is the primary spanwise wavenumber $\beta=2 \pi/L_z$), decreases by the onset of streak instabilities the energy of the spanwise-independent mode $M(\alpha,0)$ ($\alpha$ is the primary streamwise wavenumber $\alpha=2\pi/L_x$) increases as well as the energy of the wavy streak $M(\alpha,\beta)$, indicating not only the $M(0,\beta)$-to-$M(\alpha,\beta)$ energy transfer but also the $M(0,\beta)$-to-$M(\alpha,0)$ transfer via the streak instabilities. If such energy transfers were investigated through one-dimensional spanwise-Fourier-mode analysis, it would appear as an inverse interscale energy transfer from $\lambda_z=L_z$ to $\lambda_z=\infty$, since only the $M(0,\beta)$-to-$M(\alpha,0)$ transfer is observed. On the other hand, in terms of the streamwise Fourier mode, both the $M(0,\beta)$-to-$M(\alpha,\beta)$ and $M(0,\beta)$-to-$M(\alpha,0)$ energy transfers are observed as a forward energy transfer from $\lambda_x=\infty$ to $\lambda_x=L_x$. This means that the interscale energy transfer associated with the streak instabilities of the self-sustaining cycle may be observed as an inverse interscale energy transfer when observed through one-dimensional spanwise-Fourier-mode analysis. \cite{lee_2019} performed two-dimensional spectral analysis on the turbulent energy transport in channel flows at high Reynolds numbers, and observed that there exist significant energy transfers between different Fourier modes with approximately same wavenumber magnitudes but different orientations. Such `scale transfer in orientation' is observed as interscale energy transfer in different directions (i.e.~either a forward or reversed cascade) when observed through one-dimensional Fourier mode analysis, depending on if the analysis is based on the streamwise or spanwise Fourier modes. These observations in earlier studies indicate that the inverse interscale energy transfer observed based on one-dimensional Fourier mode analysis does not always mean that energy is really transported from smaller to larger scales. 

Similar tendencies to the interscale energy transfer related with the streak instabilities described above are actually found in the distributions of $tr^x_{uu}$ and $tr^z_{uu}$ presented in figures~\ref{fig:trxuu}(a) and \ref{fig:trzuu}(a), respectively. One can see that the spanwise interscale energy transfer $tr^z_{uu}$ provided by the reference case in figure~\ref{fig:trzuu}(a) indicates an inverse energy transfer from smaller to larger $\lambda_z$ particularly in the near-wall region, while the streamwise interscale transport $tr^x_{uu}$ given in figure~\ref{fig:trxuu}(a) shows only forward transfer throughout the channel. These tendencies are retained also in the minimal domains; $tr^x_{uu}$ obtained in the streamwise-minimal case indicates a forward energy transfer from the $x$-independent mode to the $\lambda_x^+<400$ range (see figure~\ref{fig:trxuu}(a)), while $tr^z_{uu}$ obtained in the spanwise-minimal case presents a reversed energy transfer from $\lambda_z=L_z$ to $z$-independent mode (figure~\ref{fig:trzuu}(a)). Such behaviours of interscale energy transfers $tr^x_{uu}$ and $tr^z_{uu}$ are similar to the above-described energy exchange between the $x$- and $z$-independent Fourier modes in the streak instabilities of the self-sustaining cycle.

\begin{figure}
\begin{center}
	\includegraphics[width=0.9\hsize]{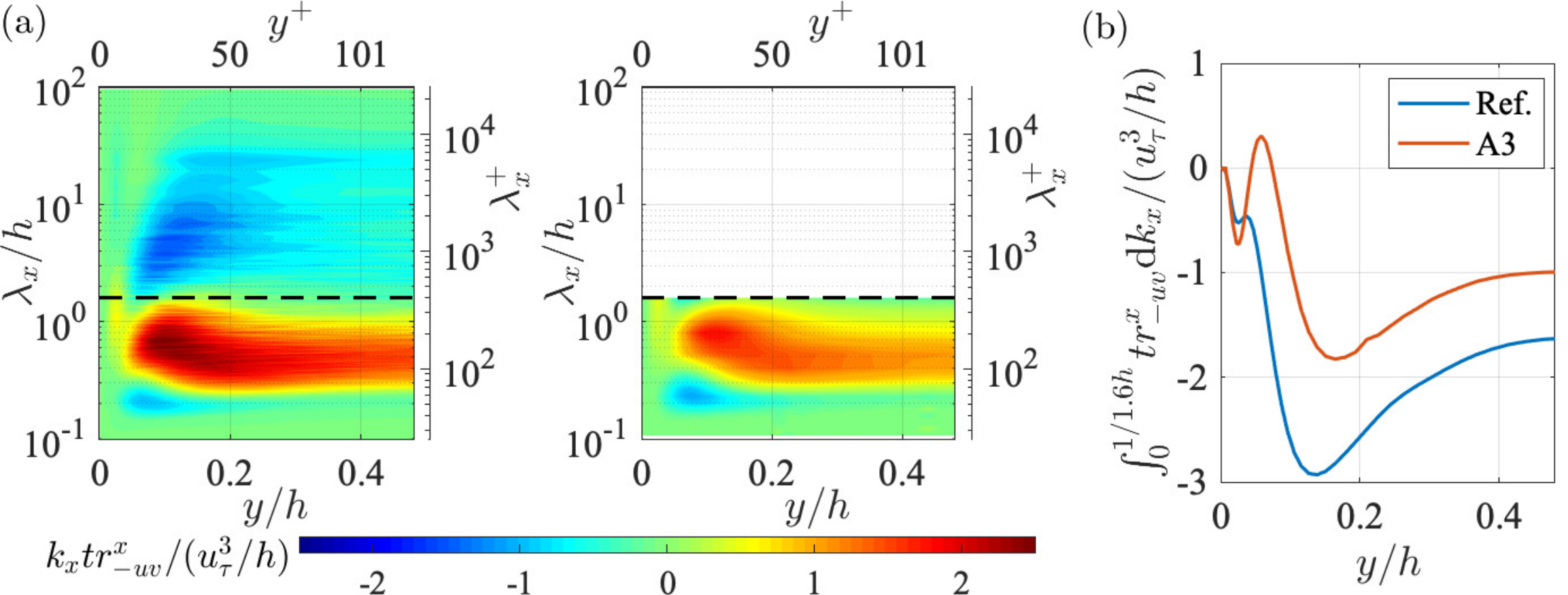}
	\caption{Distributions of the interscale transport of the Reynolds shear stress in the streamwise wavenumber direction $k_x tr^x_{-uv}$ obtained in the reference and streamwise-minimal (A3) cases, presented in the same manner as in figure~\ref{fig:exuu}(a): $y$-$\lambda_x$ diagrams in premultiplied form $k_x tr^x_{-uv}$; (b)~profiles of $tr^x_{-uv}$ integrated for $\lambda_x/h\geqq1.6$ ($0<k_x/h<2\pi/1.6$). Note that, in Case~A3, $\int^{2\pi/1.6h}_0 pr^x_{-uv}(k_x) \mathrm{d} k_x=tr^x_{-uv}(0) \Delta k_x$.}
	\label{fig:trxuv}
\end{center}
\end{figure}

Similar tendencies are also found for the interscale transport of the Reynolds shear stress. As can be seen in figure~\ref{fig:trxuv}(a), the distribution of the interscale transport of the Reynolds shear stress in the streamwise wavenumber direction $tr^x_{-uv}$ obtained in the reference case clearly presents forward transfers from larger to smaller $\lambda_x$ ranges throughout the channel, despite the spanwise transport $tr^z_{-uv}$ presenting inverse interscale transfers (see figure~\ref{fig:trzuv}(a)). The $y$-$\lambda_x$ distribution of $tr^x_{-uv}$ provided by the streamwise-minimal case also reproduces fairly well the reference-case result for the relatively small wavelength range  $\lambda_x^+<400$, and the Reynolds shear stress transferred from the $x$-independent mode, i.e.~$tr^x_{-uv}(y,0)$, in the streamwise-minimal case is equivalent to the amount of the Reynolds shear stress removed from the corresponding $\lambda_x$ range in the reference case $\lambda_x/h>1.6$, as presented in figure~\ref{fig:trxuv}(b). These behaviours of the interscale transports of the Reynolds shear stress $tr^x_{-uv}$ and $tr^z_{-uv}$ are also qualitatively similar to the above-described energy exchange between the $x$- and $z$-independent mode observed in the self-sustaining cycle by \cite{hamilton_1995}.

\begin{figure}
\begin{center}
	\includegraphics[width=0.7\hsize]{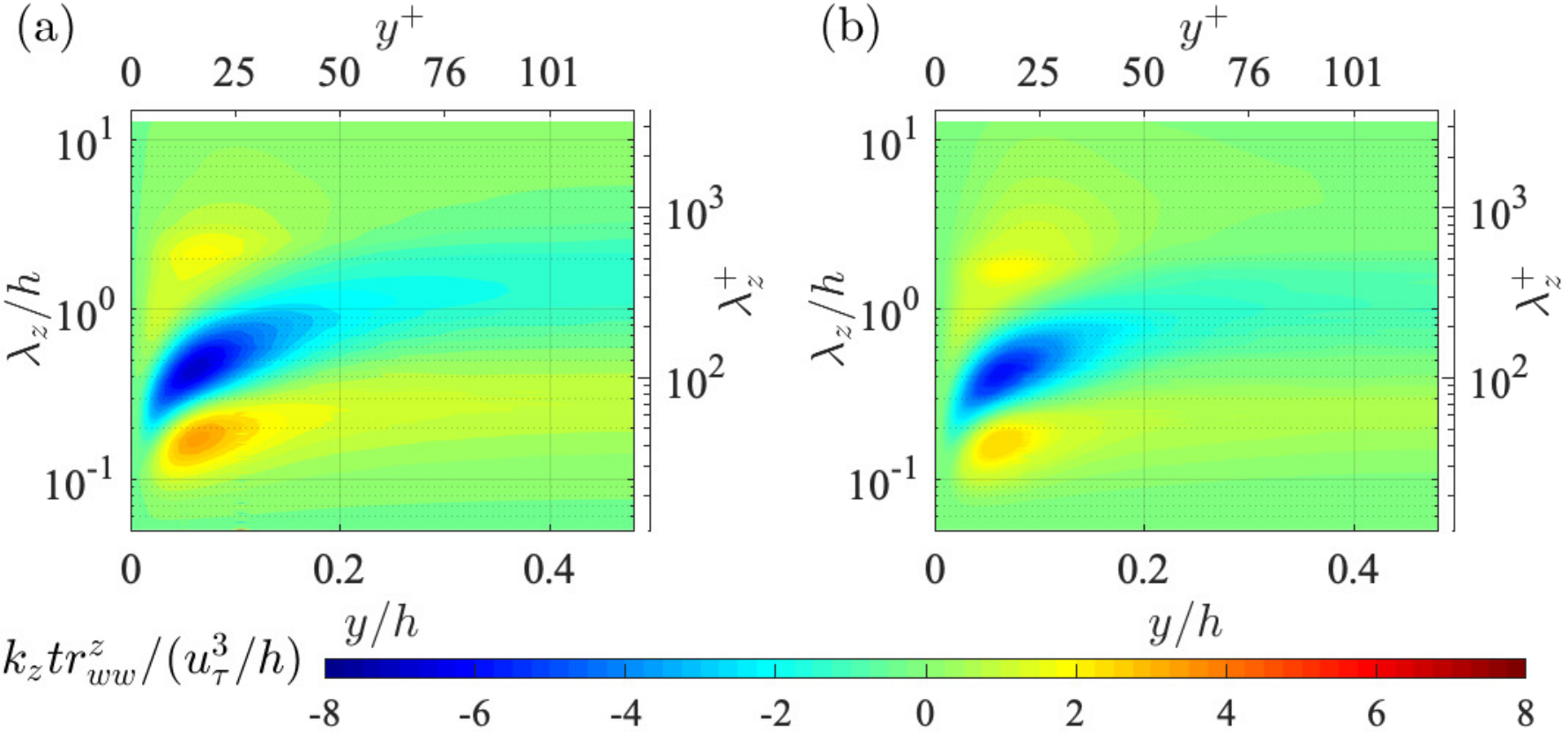}
	\caption{Space-wavelength ($y$-$\lambda_z$) diagrams of the premultiplied interscale transport of the spanwise turbulent energy in the spanwise wavenumber direction $k_z tr^z_{ww}$ obtained in (a) the reference case and (b) Case A3 (the streamwise-minimal case). The values are scaled by $u_\tau^3/h$.}
	\label{fig:trzww} 
\end{center}
\end{figure}

It should be emphasised here that the inverse interscale transfer of the spanwise turbulence intensity $\ave{w^2}$, which we suggested in Sec.~4.2 to represent the energy transfers related with the regeneration process of streamwise vortices of the self-sustaining cycle, shows qualitatively different tendencies from the energy transfer by the streak instabilities described above. Figure~\ref{fig:trzww} represents the distributions of the interscale transport of $\ave{w^2}$ in the spanwise wavenumber direction $tr^z_{ww}$ obtained in the reference and streamwise-minimal cases, and one can see here that both results clearly indicate an energy transfer from smaller to larger wavelengths in the near-wall region, similarly to the transfer in the streamwise wavenumber direction $tr^x_{ww}$ presented in figure~\ref{fig:trxww}(a). Thus, the spanwise turbulent energy $\ave{w^2}$ is transferred from smaller to larger scales in both the streamwise and spanwise wavenumber directions, unlike the above-described energy exchange between $x$- and $z$-independent modes in the streak instabilities. It also would be worth pointing out that in the $tr^x_{-uv}$ distribution (see figure~\ref{fig:trxuv}(a)) one can find a weak reversed interscale transport from smaller to larger $\lambda_x$ in the small wavelength range $\lambda_x^+<100$ throughout the channel, as well as the significant forward interscale transfers in the larger $\lambda_x$ range. Hence, the Reynolds shear stress is also transferred from smaller to larger scales in both the streamwise and spanwise wavenumber directions throughout the channel at relatively small scales. While we conjecture that the inverse $\ave{w^2}$ transfer is closely related with the generation process of streamwise vortices in the self-sustaining cycle as discussed in previous section, it is not clear what the inverse $-\ave{uv}$ transfers at small scales represents. This will be focused on in our feature studies.

\section{Conclusion}
In the present study, we have investigated the spectral transport of the Reynolds stresses in turbulent plane Couette flow at a relatively low Reynolds number $Re_\tau \approx 126$ with the streamwise- and spanwise-minimal domains, in order to illuminate the role of scale interactions in the streamwise and spanwise direction by limiting the degree of freedom in each direction. The streamwise-minimal domain has been found to, as reported in earlier studies~\citep{abe_2018}, reproduce the flow structure of the plane Couette turbulence successfully capturing statistical features of both the smaller-scale structure near the wall and the very-large-scale structure in the channel-core region, despite the substantially limited degree of freedom in the streamwise direction. In this domain, the flow field is basically separated to $x$-independent structures and the rest: the former is $x$-independent streaks that account for most of the streamwise turbulent energy $\ave{u^2}$ while the latter is the structures smaller than the streamwise-minimal length ($\lambda_x^+<400$) which account for most of the lateral turbulent energy $\ave{v^2}$ and $\ave{w^2}$. The spectral energy transport in the streamwise-minimal domain mainly consists of the turbulent energy production at $k_x=0$, interscale energy transfer from $k_x=0$ ($\lambda_x=\infty$) to $k_x>0$ ($\lambda_x^+<400$) and the energy redistribution from the streamwise to lateral turbulent energies in $\lambda_x^+<400$. The good agreement between the streamwise-minimal and reference cases indicates that such interplays between streamwise-elongated streaks and small-scale structures are the essential interactions of the streamwise length scales for both the near-wall and very-large-scale structures. It has been further revealed that the energy production at larger scales is maintained by inverse interscale transport of spanwise turbulent energy. Based on the resemblance between the closed-loop transport of the Reynolds stresses and the scenario of the self-sustaining cycle of coherent structure, we conjecture that these spectral budgets of the Reynolds stress transport mainly represent the subprocesses of the self-sustaining process of each inner and outer structure. Detailed investigations on the instantaneous flow structures support this conjecture. This indicates that the forward and reversed interscale energy transfers observed by the present spectral analysis may correspond to the streak instabilities and the regeneration of streamwise vortices of the self-sustaining cycle of the coherent structures, respectively.

In the spanwise-minimal domain, the interscale Reynolds-stress transports in the spanwise wavenumber direction $tr^z_{ij}$ indicate basically the same tendencies as observed in a very large domain, including the inverse interscale transport of the Reynolds shear stress observed by \cite{kawata_2018}, in spite of the fact that the very-large-scale structure in the channel core region does not exist in this case due to the insufficient spanwise domain width. This suggests that the interscale transport of the Reynolds stress $tr^z_{ij}$ does not represent the interactions between the very-large-scale and near-wall structures. \cite{hamilton_1995} indicated that the streak instabilities in the self-sustaining cycle involve energy transfers from the $x$- to $z$-independent mode, which would, if observed through one-dimensional spanwise Fourier mode analysis, result in a similar behaviours of $tr^z_{ij}$ as observed in the present study. 

Thus, the present study has shown that both the Reynolds-stress interscale transports in the streamwise and spanwise wavenumber directions are likely related to individual dynamics of each inner and outer structure, rather than represent their interactions. However, there have been accumulated evidences of the inner-outer interactions, such as the amplitude modulations~\citep[see the summary by][and references therein]{dogan_2019}, and it is still unclear how such interactions are reflected in the Reynolds stress transport. In order to further address this point, more detailed investigations on the interscale transfers would be needed through, for example, two-dimensional spectral analysis as done by \cite{lee_2019} or decomposing the interscale transfers into each triad interaction between Fourier modes as done by \cite{cho_2018}. Given that the inner and outer structures are located at different wall-normal positions, the spatial energy transport caused by nonlinear scale interaction, i.e.~turbulent spatial transport $d^{t,x}_{ij}$ and $d^{t,z}_{ij}$ in the spectral transport equations~(\ref{eq:exij}) and (\ref{eq:ezij}), should be focused on as well as the interscale energy transport effects. Besides, since the budget equations represent only the local gain/sink of the energy, it also would be important to investigate the flux of energy both in scale and in space, as done by \cite{cimarelli_2013}, for better understanding of dynamics in wall turbulence.

\section*{Acknowledgements}
The authors are grateful to G.~Fujimoto for his contributions to carrying out the DNSs of the present study. This work was supported by the Japan Society for the Promotion of Science (JSPS) through JSPS KAKENHI, Grant Numbers JP17J04115 and JP20K14654. The numerical simulations in the present study were performed by SX-ACE supercomputers at the Cybermedia Center of Osaka University.

\section*{Declaration of Interests}

The authors report no conflict of interest.

\appendix
\section{Derivation of the transport equations of the Reynolds stress spectra (\ref{eq:exij}) and (\ref{eq:ezij})}
\label{app:a}

For the spectral analysis of the Reynolds-stress transport budget, we use the same formulation as derived by \cite{kawata_2018,kawata_2019}. First we consider such a decomposition of the fluctuating velocities:
\begin{subeqnarray} \label{eq:dec}
	\quad \quad u_i &=& u^L_i + u^S_i,   \\
	\ave{u^L_i u^S_j} &=& \ave{u^S_i u^L_j} = 0, 
\end{subeqnarray}
where $u^L_i$ and $u^S_i$ are the large- and small-scale parts of the fluctuating velocities, respectively, and the second equation~(\ref{eq:dec}b) means that the cross correlation between the large- and small-scale parts is zero for any combination of velocity component. Such decomposition is possible with a spatial filtering based on a orthogonal mode decomposition, such as the Fourier-mode decomposition or the proper orthogonal decomposition. With such a decomposition the Reynolds stresses $\ave{u_i u_j}$ is simply decomposed into their large- and small-scale parts as
\begin{equation}
	\ave{u_i u_j} = \ave{u^L_i u^L_j} + \ave{u^S_i u^S_j}, \label{eq:dec_rs}
\end{equation}
due to (\ref{eq:dec}b). 

In this study, we employ both the transport equations of the streamwise and spanwise one-dimensional Reynolds stress spectra (\ref{eq:exij}) and (\ref{eq:ezij}) for data analysis, and for deriving these transport equations the above decompositions of the fluctuating velocities and the Reynolds stresses are done by a spatial filtering based on the streamwise and spanwise Fourier mode, respectively. Taking the derivation of the transport equation of $E^x_{ij}$ (\ref{eq:exij}) as an example, the decompositions (\ref{eq:dec}) and (\ref{eq:dec_rs}) are done by the spatial filtering based on the streamwise Fourier mode with the cutoff wavenumber $k_x$, and the transport equations of $\ave{u^L_i u^L_j}$ and $\ave{u^S_i u^S_j}$ are derived by a similar procedure to derive the transport equation of the `full' Reynolds stress (\ref{eq:rs}) as
\begin{subeqnarray} \label{eq:drs}
	\frac{\mathrm{D} \ave{u^L_i u^L_j}}{\mathrm{D} t} &=& P^{L,x}_{ij} - \varepsilon^{L,x}_{ij} + \Pi^{L,x}_{ij} + D^{p,L,x}_{ij} + D^{\nu,L,x}_{ij} + D^{t,L,x}_{ij} - Tr^x_{ij}, \\
	\frac{\mathrm{D} \ave{u^S_i u^S_j}}{\mathrm{D} t} &=& P^{S,x}_{ij} - \varepsilon^{S,x}_{ij} + \Pi^{S,x}_{ij} + D^{p,S,x}_{ij} + D^{\nu,S,x}_{ij} + D^{t,S,x}_{ij} + Tr^x_{ij}. 
\end{subeqnarray}
The details of derivation of (\ref{eq:drs}a) and (\ref{eq:drs}b) are found in \cite{kawata_2019}. The terms on the right-hand-side of (\ref{eq:drs}a) are the large-scale part of the corresponding terms of (\ref{eq:rs}), which are respectively defined as
\begin{eqnarray}
	P^{L,x}_{ij} &=& - \ave{u^L_i u^L_k} \pd{U_j}{x_k} - \ave{u^L_j u^L_k} \pd{U_i}{x_k}, \quad \varepsilon^{L,x}_{ij} = 2 \nu \ave{\pd{u^L_i}{x_k} \pd{u^L_j}{x_k}} \nonumber \\
	\Pi^{L,x}_{ij} &=& - \frac{1}{\rho} \left( \ave{p^L \pd{u^L_i}{x_j}} + \ave{ p^L \pd{u^L_j}{x_i}} \right), \quad
	D^{\nu,L,x}_{ij} = \nu \pd{^2\ave{u^L_i u^L_j}}{x_k^2}, \nonumber \\
	D^{p,L,x}_{ij} &=& -\pd{}{x_k} \left( \ave{u^L_i p^L} \delta_{ik} + \ave{u^L_j p^L} \delta_{jk} \right). \nonumber 
\end{eqnarray}
Here $\delta_{ij}$ is the Kronecker delta. The small-scale counterparts of the terms on the right-hand side of (\ref{eq:drs}b) are those with the superscript $L$ interchanged by $S$. The terms related with nonlinear interaction between the large- and small-scale parts of the velocity field are the turbulent spatial transport terms
\begin{subeqnarray}
	D^{t,L,x}_{ij}&=&-\pd{}{x_k} \left( \ave{u^L_i u^L_j u^L_k} + \ave{u^L_i u^L_j u^S_k}  
		+ \ave{u^S_i u^L_j u^S_k} + \ave{u^L_i u^S_j u^S_k} \right), \\
	D^{t,S,x}_{ij}&=&-\pd{}{x_k} \left( \ave{u^S_i u^S_j u^S_k} + \ave{u^S_i u^S_j u^L_k}
		+ \ave{u^L_i u^S_j u^L_k} + \ave{u^S_i u^L_j u^L_k} \right), 
\end{subeqnarray}
and the turbulent interscale transport term:
\begin{eqnarray}
	Tr^x_{ij}&=& -\ave{u^S_i u^S_k \pd{u^L_j}{x_k}}-\ave{u^S_j u^S_k \pd{u^L_i}{x_k}}
		+ \ave{u^L_i u^L_k \pd{u^S_j}{x_k}} + \ave{u^L_j u^L_k \pd{u^S_i}{x_k}}. \label{eq:tr}
\end{eqnarray}
It can be easily seen that the sum of the transport equations~(\ref{eq:drs}a) and (\ref{eq:drs}b) yields the classical Reynolds-stress transport equation~(\ref{eq:rs}), and the additional term $Tr^x_{ij}$ clearly represents the Reynolds stress flux from the $\ave{u^L_i u^L_j}$ to $\ave{u^S_i u^S_j}$ side across $k_x$. The streamwise one-dimensional spectra of the Reynolds stress $E^x_{ij}$ are related with the decomposed Reynolds stresses as
\begin{eqnarray}
E^x_{ij} = \pd{\ave{u^L_i u^L_j}}{k_x} = -\pd{\ave{u^S_i u^S_j}}{k_x}
\end{eqnarray}
in the wavenumber range $k_x > 0$. Hence, the $E^x_{ij}$ transport equation~(\ref{eq:exij}):
\begin{eqnarray}
\left( \pd{}{t} + U_k \pd{}{x_k} \right) E^x_{ij} = pr^x_{ij} - \epsilon^x_{ij} + \pi^x_{ij} + d^{p,x}_{ij} + d^{\nu,x}_{ij} + d^{t,x}_{ij} + tr^x_{ij}, \nonumber
\end{eqnarray}
 for the wavenumber range $k_x>0$ is derived by differentiating (\ref{eq:drs}a) with respect to $k_x$, and the terms on the right-hand side are
the $k_x$ derivative of the corresponding terms of (\ref{eq:drs}a). As for evaluating these quantities based on the discrete DNS data, the Reynolds stress spectra at wavenumber $k_{x,m}=2 \pi m /L_x$ ($m=1$, 2, $3,\cdots$) are evaluated as 
\begin{eqnarray}
E^x_{ij}(y,k_{x,m}) = \frac{\ave{u^L_i u^L_j}(y,k_{x,m}) - \ave{u^L_i u^L_j}(y,k_{x,m-1})}{\Delta k_x},
\end{eqnarray}
where $\Delta k_x = 2 \pi/L_x$, and the terms on the right-hand side of the spectrum transport equation (\ref{eq:exij}) are also obtained by the same finite differentiation. It should be noted here that $E^x_{ij}$ defined in such a way is a one-sided spectrum and only defined for $k_x>0$. The spectrum density at $k_x=0$ is separately defined as $E^x_{ij}(y,0) = \ave{u^L_i u^L_j}(y,0)/\Delta k_x$, so that it satisfies 
\begin{eqnarray}
\ave{u_i u_j}(y) = \sum^\infty_{m=0} E^x_{ij} (y,k_{x,m}) \Delta k_x.  
\end{eqnarray}
Corresponding to this the transport equation of $E^x_{ij}(y,0)$ is obtained by dividing the equation (\ref{eq:drs}a) at $k_x=0$ by $\Delta k_x = 2 \pi/L_x$. Therefore, the terms on the right-hand side of the $E^x_{ij}(y,0)$ equation are accordingly defined as $pr^x_{ij}(y,0)=P^{L,x}_{ij}(y,0)/\Delta k_x$, $\epsilon^x_{ij}(y,0)=\varepsilon^{L,x}_{ij}(y,0)/\Delta k_x$, $\pi^x_{ij}(y,0)=\Pi^{L,x}_{ij}(y,0)/\Delta k_x$, $\cdots$, and $tr^x_{ij}(y,0)=-Tr^x_{ij}(y,0)/\Delta k_x$. The transport equations of the spanwise one-dimensional spectra $E^z_{ij}$, the equation~(\ref{eq:ezij}), are also derived by the same manner, with the spatial filtering based on the spanwise Fourier mode.

\section{Interscale transport on the $x$-independent mode }
\label{app:b}
As explained in the previous section, the interscale transport term in the transport equation of the Reynolds stress spectrum at $k_x=0$ is defined as
\begin{eqnarray}
	tr^x_{ij}(y,0) &=& -\frac{Tr^x_{ij}(y,0)}{\Delta k_x}, \nonumber 
\end{eqnarray}
with the interscale Reynolds-stress flux $Tr^x_{ij}$ defined in (\ref{eq:tr}). At $k_x=0$, the large-scale component $u^L_i$ in $Tr^x_{ij}(y,0)$ is the fluctuating velocity spatially low-pass-filtered with the cutoff wavenumber $k_x=0$, i.e.~$u^L_i=\ave{u_i}_x$, where $\ave{}_x$ denotes spatial averaging in the $x$-direction. Hence, one can easily see that $\ave{u^S_i}_x=0$. With such $u^L_i$ and $u^S_i$, the third and fourth terms in the right-hand side of (\ref{eq:tr}) are zero. For instance, 
\begin{eqnarray}
\ave{u^L_i u^L_k \pd{u^S_j}{x_k}} = \ave{u^L_i u^L_k \pd{u^S_j}{x_k}}_{x,z,t} = \ave{u^L_i u^L_k \ave{\pd{u^S_j}{x_k}}_x}_{z,t}=0.
\end{eqnarray}
Furthermore, $\partial u^L_i /\partial x = 0$ since $u^L_i$ is independent of $x$. Hence, all terms related with $\partial u^L_i /\partial x$ are also zero. Therefore, $tr^x_{ij}(y,0)$ is simplified as
\begin{eqnarray}
tr^x_{ij}(y,0) &=& \left(\ave{u^S_i v^S \pd{u^L_j}{y}}+\ave{u^S_j v^S \pd{u^L_i}{y}} \right. \nonumber \\
&& \hspace{2cm} + \left. \ave{u^S_i w^S \pd{u^L_j}{z}}+\ave{u^S_j w^S \pd{u^L_i}{z}} \right) \frac{1}{\Delta k_x}.
\end{eqnarray}
In particular, $tr^x_{uu}(y,0)$, which is intensively discussed in Sec.~\ref{sec:ssp}, is
\begin{eqnarray}
tr^x_{uu}(y,0) = 2 \left(\ave{u^S v^S \pd{u^L}{y}}+\ave{u^S w^S \pd{w^L}{z}} \right) \frac{1}{\Delta k_x}.
\end{eqnarray}

\bibliographystyle{jfm}

\end{document}